\documentclass[10pt,onecolumn,notitlepage]{revtex4-1} 

\usepackage{amsfonts}
\usepackage{amsmath}
\usepackage{braket}

\usepackage{graphicx}

\usepackage{pdfpages}
\usepackage{pgffor}

\makeatletter
\AtBeginDocument{\let\LS@rot\@undefined}
\makeatother

\begin{document}
	
\title{Higher-order interactions in quantum optomechanics:\\Analytical solution of nonlinearity}
	
\author{Sina Khorasani}

\affiliation{School of Electrical Engineering, Sharif University of Technology, Tehran, Iran\\ \'{E}cole Polytechnique F\'{e}d\'{e}rale de Lausanne, CH-1015, Lausanne, Switzerland}
\email{sina.khorasani@epfl.ch}

\begin{abstract}
A method is described to solve the nonlinear Langevin equations arising from quadratic interactions in quantum mechanics. While, the zeroth order linearization approximation to the operators is normally used, here first and second order truncation perturbation schemes are proposed. These schemes employ higher-order system operators, and then approximate number operators with their corresponding mean boson numbers, only where needed. Spectral densities of higher-order operators are derived, and an expression for the second-order correlation function at zero time-delay has been found, which reveals that the cavity photon occupation of an ideal laser at threshold reaches $\sqrt{6}-2$, in good agreement with extensive numerical calculations. As further applications, analysis of the quantum anharmonic oscillator, calculation of $Q-$functions, analysis of quantum limited amplifiers, and nondemoliton measurements.
\end{abstract}


\maketitle

\section{Introduction}

In quantum optomechanics the standard interaction Hamiltonian is simply the product of photon number $\hat{n}=\hat{a}^\dagger\hat{a}$ and the position $x_{\textrm{zp}}(\hat{b}+\hat{b}^\dagger)$ operators \cite{1,2,2a,3,4,5}, where $x_{\textrm{zp}}$ is the zero-point motion, and $\hat{a}$ and $\hat{b}$ are respectively the photon and phonon annihilators. This type of interaction can successfully describe a vast range of phenomena, including optomechanical arrays \cite{6,7,8,9,10,11,12}, squeezing of phonon states \cite{14,14a,14b}, non-reciprocal optomechanics \cite{15,16,17,17a}, Heisenberg’s limited measurements \cite{17b}, sensing \cite{18,19,20}, engineered dissipation and states \cite{21,22}, and non-reciprocal acousto-optics \cite{23}. In all these applications, the mathematical toolbox to estimate the measured spectrum is Langevin equations \cite{24,25,26,26a}.

Usually, the analysis of quantum optomechanics is done within the linearized approximation of photon ladder operators, normally done as $\hat{a}\rightarrow\bar{a}+\delta\hat{a}$ with $|\bar{a}|^2=\bar{n}$ being the mean cavity photon number, while nonlinear terms in $\delta\hat{a}$ are ignored. But this suffers from limited accuracy wherever the basic optomechanical interaction $\mathbb{H}_{\rm OM}=\hbar g_0\hat{n}(\hat{b}+\hat{b}^\dagger)$ is either vanishingly small or non-existent. In fact, the single-photon interaction rate $g_0$ can be identically made zero by appropriate design \cite{27,28,29,30}, when quadratic or even quartic effects are primarily pursued. This urges need for accurate knowledge of higher-order interaction terms. 

Some other optomechanical phenomena such as four-wave mixing, also can be suitably understood by incorporation of higher-order interaction terms \cite{31}. Recent experiments \cite{32,33} have already established the significance and prominent role of such type of nonlinear interactions. In fact, quadratic nonlinear optomechanics \cite{33.0,33.1,33.2,33.3,33.4,33.5,33.6,33.6.0,33.6.1,33.6.2,33.6.3,33.6.4,33.6.5,33.6.5a,33.6.6,33.6.7} is now a well recognized subject of study even down to the single-photon level \cite{33.7}, for which circuit analogues have been constructed \cite{33.8.0,33.8} and may be regarded as fairly convenient simulators \cite{33.8a,33.8b,33.8c} of much more complicated experimental optomechanical analogues. Dual formalisms of quadratic optomechanics are also found in ultracold atom traps \cite{33.9,33.9a} as well as optical levitation \cite{33.10}. Such types of nonlinear interactions also appear elsewhere in anharmonic quantum circuits \cite{33.10.0}. Quadratic interactions are in particular important for energy and non-demolition measurements of mechanical states \cite{1,2,3,33.11,33.12,33.13}. While the simple linearization of operators could be still good enough to explain some of the observations, there remains a need for an exact and relatively simple mathematical treatment. Method of Langevin equations also normally fails, and other known methods such as expansion unto number states and master equation, require lots of computation while giving little insight to the problem. 

Perturbative expansions and higher-order operators have been used by other researchers to study noise spectra of lasers \cite{33.15,33.16,33.19,33.20}. Also, the master equation approach \cite{33a,33b} can be used in combination with the quasi-probablity Wigner functions \cite{33c,33d} to yield integrable classical Langevin equations. Nevertheless, a method recently has been proposed \cite{34}, which offers a truncation correlation scheme for solution of driven-dissipative multi-mode systems. While being general, it deals with the time evolution of expectation values instead of operators within the truncation accuracy, so the corresponding Langevin equations cannot be analytically integrated. 

Alternatively, a first-order perturbation has been proposed to tackle the nonlinear quadratic optomechanics \cite{34a}. This method perturbatively expands the unknown parameters of classical Langevin equations for the nonlinear system, and proceeds to the truncation at first order. However, the expansion is accurate only where the ratio of photon loss rate to mechanical frequency $\kappa/\Omega$ is large. This condition is strongly violated for instance in superconductive electromechanical systems. 

So far and to the best knowledge of author, no treatment of quadratic interactions using Langevin equations for operators has been reported. This paper presents a perturbative mathematical treatment within the first and second order approximations to the nonlinear system of Langevin equations, which ultimately result in an integrable system of quantum mechanical operators. The trick here is to introduce operators of higher dimensionality into the solution space of the problem. Having their commutators calculated, it would be possible to set up an extended system of Langevin equations which could be conveniently solved by truncation at the desirable order. To understand how it works, one may consider the infamous first order quadratic nonlinear Riccati differential equation \cite{34b,34c}, which is exactly integrable if appropriately transformed as a system of two coupled linear first order differential equations. Alternatively, Riccati equation could be exactly transformed into a linear second order differential equation, too. But this is not what we consider here, since it will result in a much more complicated second-order system of Langevin equations involving derivatives of noise terms. 

The method introduced here is useful in other areas of quantum physics \cite{33.9,33.10} than optomechanics, where nonlinearities such as anharmonic or Kerr interactions are involved. We also describe how the $Q-$functions could be obtained for the anharmonic oscillator. Further applications of nonlinear stochastic differential equations \cite{34.8,34.9,34.10} beyond stochastic optomechanics \cite{33.6.5,33.6.5a} includes finance and stock-market analysis \cite{34.11}, turbulence \cite{34.12,34.12a}, hydrology and flood prediction \cite{34.13}, and solar energy \cite{34.14}. Also, the Fokker-Planck equation \cite{33.20,34.15,34.16,34.16a,34.16b} is actually equivalent to the nonlinear Schr\"{o}dinger equation with bosonic operator algebra, and its moments \cite{34.17} translate into nonlinear Langevin equations. Similarly, this method can deal with side-band generation in optomechanics \cite{35}, superconducting circuits \cite{35a}, as well as spontaneous emission in open systems \cite{36,36.0}. Applications in estimation of other parameters such as the second order correlation $g^{(2)}(0)$ \cite{36a,36b,36c,36d}, quantum limited amplifiers \cite{36e,36f} and quantum nondemolition measurements \cite{36f,36g,36h,36i} are demonstrated, and furthermore it is found that an unsqueezed ideal laser reaches $\sqrt{6}-2$ cavity photons at threshold.

\section{Theory}

\subsection{Hamiltonian}

A nonlinear quadratic optomechanical interaction in the most general form \cite{37} is here defined as 
\begin{equation}
\label{eq1}
\mathbb{H}=\hbar \gamma (\hat{b}\pm\hat{b}^\dagger)^2(\hat{a}\pm\hat{a}^\dagger)^2,
\end{equation}
\noindent
where $\gamma$ is the interaction rate. Furthermore, bosonic photon $\hat{a}$ and phonon $\hat{b}$ ladder operators satisfy $[\hat{b},\hat{b}^\dagger]=[\hat{a},\hat{a}^\dagger]=1$ as well as $[\hat{b},\hat{a}]=[\hat{b},\hat{a}^\dagger]=0$. Meanwhile, quadratic interactions normally are \cite{1,2,3}
\begin{equation}
\label{eq2}
\mathbb{H}=\hbar \gamma \hat{a}^\dagger\hat{a}(\hat{b}\pm\hat{b}^\dagger)^2,
\end{equation}
\noindent
which by defining the photon number operator $\hat{n}=\hat{a}^\dagger\hat{a}$ takes essentially the same algebraic form.

Direct expansion of (\ref{eq1}) shows that it essentially brings in a different interaction type compared to (\ref{eq2}). Doing so, we obtain $\mathbb{H}=\hbar \gamma (\hat{b}^2+\hat{b}^{\dagger2}\pm 2\hat{m}\pm 1)^2(\hat{a}^2+\hat{a}^{\dagger2}\pm 2\hat{n}\pm 1)$ where $\hat{m}=\hat{b}^\dagger\hat{b}$. Hence, (\ref{eq1}) includes interactions of type $\hat{a}^2\hat{b}^2$, $\hat{a}^2\hat{b}^{\dagger2}$, and so on, which are absent in (\ref{eq2}). It should be noticed that the widely used standard optomechanical interaction $\mathbb{H}_{\rm OM}$ results in nonlinear and linear Langevin equations when expressed respectively in the terms of $\{\hat{a},\hat{b}\}$ and $\{\hat{n},\hat{x}\}$. Hence, this type of interaction is not addressed here. In addition to the above Hamiltonians (\ref{eq1},\ref{eq2}), there exist still other types of nonlinear optomechanical interactions \cite{14b,38} such as $\mathbb{H}=\hbar g (\hat{b}\pm\hat{b}^\dagger)(\hat{a}^2\pm\hat{a}^{\dagger2})$, which is also not considered explicitly here, but can be well treated using the scheme presented in this article.  

\subsection{Linear Perturbation}

This approach is being mostly used by authors to solve the systems based on either (\ref{eq1}) or (\ref{eq2}). To this end, ladder field operators are replaced with their perturbations, while product terms beyond are neglected and truncated. Obviously, this will give rise to interactions of the type $\hbar(\hat{b}\pm\hat{b}^\dagger)^2(q\delta\hat{a}+q^\ast\delta\hat{a}^\dagger)$, where $q=2\gamma(\bar{a}\pm\bar{a}^\ast)$ for (\ref{eq1}) and $q=\gamma\bar{a}$ for (\ref{eq2}) is some complex constant in general, and $\delta\hat{a}$ now represents the perturbation term around the steady state average $|\bar{a}|=\sqrt{\bar{n}}$. This technique is mostly being referred to as the linearization of operators, and directly leads to an integrable set of Langevin equations if also applied to the mechanical displacement as well.

\subsection{Square Field Operators}

Here, we define the square field operators \cite{37}
\begin{eqnarray}
\label{eq3}
\hat{c}&=&\frac{1}{2}\hat{a}^2,\\ \nonumber
\hat{d}&=&\frac{1}{2}\hat{b}^2.
\end{eqnarray}
\noindent
for photons, which obviously satisfy $[\hat{c},\hat{a}]=[\hat{c},\hat{b}]=[\hat{d},\hat{a}]=[\hat{d},\hat{b}]=[\hat{c},\hat{d}]=0$. Now, it is not difficult to verify that these operators furthermore satisfy the commutation relationships
\begin{eqnarray}
\label{eq4}
[\hat{c},\hat{c}^\dagger]&=&\hat{n}+\frac{1}{2},\\ \nonumber
[\hat{c},\hat{n}]&=&2\hat{c},\\ \nonumber
[\hat{c}^\dagger,\hat{n}]&=&-2\hat{c}^\dagger,\\ \nonumber
[\hat{c},\hat{a}^\dagger]&=&\hat{a}.
\end{eqnarray}
Defining the phonon number operator as $\hat{m}=\hat{b}^\dagger\hat{b}$, in a similar manner we could write
\begin{eqnarray}
\label{eq5}
[\hat{d},\hat{d}^\dagger]&=&\hat{m}+\frac{1}{2},\\ \nonumber
[\hat{d},\hat{m}]&=&2\hat{d},\\ \nonumber
[\hat{d}^\dagger,\hat{m}]&=&-2\hat{d}^\dagger,\\ \nonumber
[\hat{d},\hat{b}^\dagger]&=&\hat{b}.
\end{eqnarray}

The set of commutator equations (\ref{eq4}) and (\ref{eq5}) enables us to treat the quadratic nonlinear interaction perturbatively to the desirable accuracy, as is described in the following.

\subsection{Langevin Equations}

The input/output formalism \cite{24,25,26,26a} can be used to assign decay channels to each of the quantum variables of the system. This will result in the set of Langevin equations  
\begin{equation}
\label{eq6}
\frac{d}{dt}\{A\}=[\textbf{M}]\{A\}-\sqrt{[\Gamma]}\{A_{\rm in}\},
\end{equation}
\noindent
where $\{A\}$ is the system vector, $[\textbf{M}]$ is the coefficients matrix whose eigenvalues need to have negative or vanishing real parts to guarantee stability, and $[\Gamma]$ is a real-valued matrix which is diagonal if all noise terms corresponding to the members of $\{A\}$ are mutually independent. When $[\textbf{M}]$ is independent of $\{A\}$, (\ref{eq6}) is linear and integrable and otherwise nonlinear and non-integrable. If  $[\textbf{M}(t)]$ is a function of time, then (\ref{eq6}) is said to be time-dependent. Furthermore, $\{A_{\rm in}\}$ represents the input fields to the system at the respective ports, and $\{A_{\rm out}\}$ is the output fields, which are related together as \cite{4,5,6}
\begin{equation}
\label{eq7}
\{A_{\rm out}\}=\{A_{\rm in}\}+\sqrt{[\Gamma]}\{A\}.
\end{equation}
\noindent
Here, $[\Gamma]$ is supposed to be diagonal for simplicity. From the scattering matrix formalism we also have
\begin{equation}
\label{eq8}
\{A_{\rm out}\}=[\textbf{S}]\{A_{\rm in}\}.
\end{equation}
\noindent
Hence, taking $w$ as the angular frequency and performing a Fourier transform on (\ref{eq6}), the scattering matrix is found by using (\ref{eq7}) and (\ref{eq8}) as
\begin{equation}
\label{eq9}
[\textbf{S}(w)]=[\textbf{I}]-\sqrt{[\Gamma]}\left(iw[\textbf{I}]+[\textbf{M}]\right)^{-1}\sqrt{[\Gamma]}.
\end{equation}
\noindent
Hence,  $[\textbf{S}]$ is well-defined if $[\textbf{M}]$ is known. This can be obtained by using the Langevin equations 
\begin{equation}
\label{eq10}
\dot{\hat{z}}=\frac{d}{dt}\hat{z}=-\frac{i}{\hbar}[\hat{z},\mathbb{H}]-[\hat{z},\hat{x}^\dagger](\frac{1}{2}\Gamma\hat{x}+\sqrt{\Gamma}\hat{z}_{\rm in})+(\frac{1}{2}\Gamma\hat{x}^\dagger+\sqrt{\Gamma}\hat{z}_{\rm in}^\dagger)[\hat{z},\hat{x}],
\end{equation}
\noindent
where $\hat{x}$ is any system operator, which is here taken to be the same as $\hat{z}$ to comply with (\ref{eq8}).

By setting either $\hat{z}=\hat{c}$ or $\hat{z}=\hat{d}$ the commutators in (\ref{eq10}) by (\ref{eq4}) or (\ref{eq5}) always lead back to the same linear combination of these forms. Thus, the new set of Langevin equations is actually linear in terms of the square or higher-order operators, if perturbatively truncated at a finite order. So, instead of solving the nonlinear system in linearized $2\times 2$ space $\{A\}^{\rm T}=\{\hat{a},\hat{b}\}$, one may employ an expanded dimensional space with increased accuracy. There, truncation and sometimes mean field approximations are necessary to restrict the dimension, since commutators of new operators mostly lead to even higher-orders and are thus not closed under commutation. As examples, a $4\times 4$ space $\{A\}^{\rm T}=\{\hat{a},\hat{d},\hat{d}^\dagger,\hat{m}\}$ truncated at the first-order, or a $6\times 6$ space $\{A\}^{\rm T}=\{\hat{c},\hat{c}^\dagger,\hat{n},\hat{d},\hat{d}^\dagger,\hat{m}\}$ truncated at the second-order could be used for (\ref{eq1},\ref{eq2}). To illustrate the application of this method, we describe two examples in the next section. It could be extended to the accuracy of the second-order perturbation too, by defining appropriate cross product operator terms between photonic and phononic partitions.   

\section{Examples}

Here, we describe two examples from the nonlinear interactions of having type (\ref{eq1}) or (\ref{eq2}).

\subsection{Standard Quadratic Interaction (\ref{eq2})}

Analysis of such systems requires analysis in a 4-dimensional space, spanned by $\{A\}^{\rm T}=\{\hat{a},\hat{d},\hat{d}^\dagger,\hat{m}\}$. Taking the plus sign here without loss of generality and after dropping a trivial non-interacting term $\mathbb{H}_0=\hbar\gamma\hat{n}$, the nonlinear interaction  is
\begin{equation}
\label{eq11}
\mathbb{H}=2\hbar\gamma\hat{n}(\hat{d}+\hat{d}^\dagger+\hat{m}).
\end{equation}

This can be found by expansion of (\ref{eq2}), plugging in (\ref{eq3}) and $[\hat{b},\hat{b}^\dagger]=1$, and dropping a trivial term $\hbar\gamma\hat{n}$. Using (\ref{eq5}), $[\hat{a},\hat{n}]=\hat{a}$ and $[\hat{a}^\dagger,\hat{n}]=-\hat{a}^\dagger$ in the non-rotating frame of operators, and ignoring the self-energy Hamiltonian $\mathbb{H}_{\rm self}=\hbar(\omega+\gamma)\hat{n}+\hbar\Omega\hat{m}$ for the moment, Langevin equations become
\begin{eqnarray}
\nonumber
\dot{\hat{a}}&=&-2i\gamma\hat{a}(\hat{d}+\hat{d}^\dagger+\hat{m})-\frac{1}{2}\Gamma_1\hat{a}-\sqrt{\Gamma_1}\hat{a}_{\textrm{in}},\\ \nonumber
\dot{\hat{d}}&=&-2i\gamma\hat{n}(2\hat{d}+\hat{m}+\frac{1}{2})-(\hat{m}+\frac{1}{2})(\frac{1}{2}\Gamma_2\hat{d}+\sqrt{\Gamma_2}\hat{d}_{\textrm{in}}), \\ \nonumber
\dot{\hat{d}}^\dagger&=&2i\gamma\hat{n}(2\hat{d}^\dagger+\hat{m}+\frac{1}{2})-(\hat{m}+\frac{1}{2})(\frac{1}{2}\Gamma_2\hat{d}^\dagger+\sqrt{\Gamma_2}\hat{d}_{\textrm{in}}^\dagger),\\ \label{eq12}
\dot{\hat{m}}&=&4i\gamma\hat{n}(\hat{d}-\hat{d}^\dagger).
\end{eqnarray}

So far, the set of equations (\ref{eq12}) is exact. However, integration of (\ref{eq12}) is still not possible at this stage, and taking Fourier transformation must be done later when arriving at a linear operator system. We present a first-order and second-order perturbative method to deal with this difficulty. 

It should be furthermore noticed that using a non-rotating frame with the self-energy Hamiltonian $\mathbb{H}_{\rm self}$ not ignored, would have resulted in identical equations, except with the addition of the trivial terms $-i\Delta\hat{a}$, $-i2\Omega\hat{d}$, and $+i2\Omega\hat{d}^\dagger$ respectively to the first three equations, where $\Delta=\omega+\gamma-\nu$ is the optical detuning with $\nu$ being the cavity optical resonance frequency, and $\omega$ and $\Omega$ are respectively the optical and mechanical frequencies. Also, the damping coefficient in high mechanical quality factor $Q_{\rm m}$ limit could be estimated as $\Gamma_2=2\Gamma_{\rm m}$, where $\Gamma_{\rm m}$ is the damping rate of the $\hat{b}$ phononic field. Here, it is preferable not to use the rotating frames since the coefficients matrix $[\textbf{M}]$ becomes time-dependent.

\subsubsection{First-order Perturbation to (\ref{eq12})}

Now, if the photon and phonon baths each have a mean boson number respectively as $\left<\hat{n}\right>=\bar{n}$ and $\left<\hat{m}\right>=\bar{m}$, we could immediately write down the linear system of equations in the non-rotating frame of operators and neglection of self-energies $\mathbb{H}_{\rm self}$ as
\begin{eqnarray}
\nonumber
\dot{\hat{a}}&=&-3i\gamma\bar{m}\hat{a}-i\gamma\bar{a}\hat{d}-i\gamma\bar{a}\hat{d}^\dagger-\frac{1}{2}\Gamma_1\hat{a}-\sqrt{\Gamma_1}\hat{a}_{\textrm{in}},\\ \nonumber
\dot{\hat{d}}&=&-2i\gamma\bar{n}\left(2\hat{d}+\hat{m}+\frac{1}{2}\right)-\left(\bar{m}+\frac{1}{2}\right)\left(\frac{1}{2}\Gamma_2\hat{d}+\sqrt{\Gamma_2}\hat{d}_{\textrm{in}}\right), \\ \nonumber
\dot{\hat{d}}^\dagger&=&2i\gamma\bar{n}\left(2\hat{d}^\dagger+\hat{m}+\frac{1}{2}\right)-\left(\bar{m}+\frac{1}{2}\right)\left(\frac{1}{2}\Gamma_2\hat{d}^\dagger+\sqrt{\Gamma_2}\hat{d}_{\textrm{in}}^\dagger\right),\\ \label{eq13}
\dot{\hat{m}}&=&4i\gamma\bar{n}\left(\hat{d}-\hat{d}^\dagger\right),
\end{eqnarray}
\noindent
which is now exactly integrable. Here, we use the linearization $2\hat{a}\hat{d}=(\bar{a}+\delta\hat{a})\hat{d}+\hat{a}(\bar{d}+\delta\hat{d})\rightarrow\bar{a}\hat{d}+\bar{d}\hat{a}$, where $\bar{d}=\frac{1}{2}\bar{a}^2$ and higher-order terms of the form $\delta\hat{a}\delta\hat{d}$ are dropped, and so on. But this cannot be applied to $\hat{n}\hat{m}=\hat{a}^\dagger\hat{a}\hat{m}$ since $\hat{n}$ and $\hat{a}^\dagger$ are absent from the basis. Furthermore, any linearization of this expansion would generate terms $\hat{a}\hat{m}$ and $\hat{a}^\dagger\hat{m}$ which are still nonlinear. Both of these issues can be resolved by a second-order perturbation as follows next. This results in the operator equations
\begin{eqnarray}
\raggedleft
\nonumber
\frac{d}{dt}\begin{Bmatrix}
\hat{a}\\
\hat{d}\\
\hat{d}^\dagger\\
\hat{m}
\end{Bmatrix}&=&\begin{bmatrix}
-i3\gamma\bar{m}-\frac{1}{2}\Gamma_1 & -i\gamma\bar{a} & -i\gamma\bar{a} & 0\\
0 & -i4\gamma\bar{n}-\frac{1}{2}\left(\bar{m}+\frac{1}{2}\right)\Gamma_2 & 0 & -i2\gamma\bar{n}\\
0 & 0 & +i4\gamma\bar{n}-\frac{1}{2}\left(\bar{m}+\frac{1}{2}\right)\Gamma_2 & i2\gamma\bar{n} \\
0 & i4\gamma\bar{n} & -i4\gamma\bar{n} & 0
\end{bmatrix}\begin{Bmatrix}
\hat{a}\\
\hat{d}\\
\hat{d}^\dagger\\
\hat{m}
\end{Bmatrix}\\ \label{eq14}
&-&\begin{Bmatrix}
\sqrt{\Delta_1}\hat{a}_{\textrm{in}}\\
\sqrt{\Delta_2}\hat{d}_{\textrm{in}}\\
\sqrt{\Delta_2}\hat{d}^\dagger_{\textrm{in}}\\
0
\end{Bmatrix},
\end{eqnarray}	
\noindent
where $\sqrt{\Delta_1}=\sqrt{\Gamma_1}$ and $\sqrt{\Delta_2}=\left(\bar{m}+\frac{1}{2}\right)\sqrt{\Gamma_2}$. The set of equations (\ref{eq14}) is linear and can be easily addressed by standard methods of stochastic Langevin equations used in optomechanics \cite{1,2,3,24,25} and elsewhere. More specifically, one may employ analytical Fourier methods in frequency domain as an matrix algebraic problem to obtain spectra of variables, or integrate the system numerically by stochastic numerical methods in time domain to obtain time dependent behavior of expectation values.  

All that remains is to find the average cavity boson numbers for photons $\bar{n}$ and phonons $\bar{m}$. In order to do this, one may first arbitrate $d/dt=0$ in (\ref{eq13}) at steady state, and then use the equality of real parts in first equation to find the expression for $\bar{n}$. Doing this, results in $\bar{n}=4|\bar{a}_{\rm in}|^2/\Gamma_1$ where $|\bar{a}_{\rm in}|$ represents the amplitude of coherent laser input. Also, the initial cavity phonon occupation number at $t=0$ could be estimated simply as  $\bar{m}=1/\left[\exp(\hbar\Omega/k_{\rm B}T)-1\right]$ \cite{24,25}, where $k_{\rm B}T$ is the thermal energy with $k_{\rm B}$ and $T$ being respectively the Boltzmann's constant and absolute temperature. Detailed numerical examinations reveal that the system of equations (\ref{eq14}) is generally very well stable with $\Re\{{\rm eig}[{\bf M}]\}<0$ at sufficiently low optical intensities.

\subsection{Full Quadratic Interaction (\ref{eq1})}

Analysis of a fully quadratic system requires analysis in a $6\times 6$ dimensional space, spanned by $\{A\}^{\rm T}=\{\hat{c},\hat{c}^\dagger,\hat{n},\hat{d}, \hat{d}^\dagger,\hat{m}\}$. Taking both of the plus signs here, the Hamiltonian could be written as
\begin{equation}
\label{eq15}
\mathbb{H}=4\hbar\gamma(\hat{d}+\hat{d}^\dagger+\hat{m})(\hat{c}+\hat{c}^\dagger+\hat{n}),
\end{equation}
\noindent
where a trivial non-interacting term $\mathbb{H}_0=2\hbar\gamma(1+\hat{n}+\hat{m}+\hat{d}+\hat{c}+\hat{d}^\dagger+\hat{c}^\dagger)$ is dropped. The set of Langevin equations can be obtained in a similar manner, and in non-rotating frame of operators with neglection of self-energies $\mathbb{H}_{\rm self}=\hbar(\omega+2\gamma)\hat{n}+\hbar(\Omega+2\gamma)\hat{m}$ for the moment, results in
\begin{eqnarray}
\raggedleft
\label{eq16}
\dot{\hat{c}}&=&-i4\gamma(\hat{d}+\hat{d}^\dagger+\hat{m})\left(2\hat{c}+\hat{n}+\frac{1}{2}\right)-\left(\hat{n}+\frac{1}{2}\right)\left(\frac{1}{2}\Gamma_1\hat{c}+\sqrt{\Gamma_1}\hat{c}_{\textrm{in}}\right),\\ \nonumber
\dot{\hat{c}}^\dagger&=&i4\gamma(\hat{d}+\hat{d}^\dagger+\hat{m})\left(2\hat{c}^\dagger+\hat{n}+\frac{1}{2}\right)-\left(\hat{n}+\frac{1}{2}\right)\left(\frac{1}{2}\Gamma_1\hat{c}^\dagger+\sqrt{\Gamma_1}\hat{c}_{\textrm{in}}^\dagger\right),\\ \nonumber
\dot{\hat{n}}&=&i8\hbar\gamma(\hat{d}+\hat{d}^\dagger+\hat{m})(\hat{c}-\hat{c}^\dagger),\\ \nonumber
\dot{\hat{d}}&=&-i4\gamma(\hat{c}+\hat{c}^\dagger+\hat{n})\left(2\hat{d}+\hat{m}+\frac{1}{2}\right)-\left(\hat{m}+\frac{1}{2}\right)\left(\frac{1}{2}\Gamma_2\hat{d}+\sqrt{\Gamma_2}\hat{d}_{\textrm{in}}\right),\\ \nonumber
\dot{\hat{d}}^\dagger&=&i4\gamma(\hat{c}+\hat{c}^\dagger+\hat{n})\left(2\hat{d}^\dagger+\hat{m}+\frac{1}{2}\right)-\left(\hat{m}+\frac{1}{2}\right)\left(\frac{1}{2}\Gamma_2\hat{d}^\dagger+\sqrt{\Gamma_2}\hat{d}_{\textrm{in}}^\dagger\right),\\ \nonumber
\dot{\hat{m}}&=&i8\hbar\gamma(\hat{c}+\hat{c}^\dagger+\hat{n})(\hat{d}-\hat{d}^\dagger).
\end{eqnarray}
\noindent
Similar to (\ref{eq12}), the damping rate for sufficiently high optical quality factors $Q$ could be estimated as $\Gamma_1=2\kappa$, where $\kappa$ is the damping rate of the $\hat{a}$ photonic field.

Quite clearly, should we have not ignored the self-energy Hamiltonian $\mathbb{H}_{\rm self}$, then addition of the diagonal terms $-i2\Delta\hat{c}$, $+i2\Delta\hat{c}^\dagger$ to the first two where $\Delta=\omega+2\gamma-\nu$ with $\nu$ being the optical cavity resonance frequency, and similarly $-i2\Omega\hat{d}$ and $+i2\Omega\hat{d}^\dagger$ to the fourth and fifth equations would have been necessary. These are not shown here only for the sake of convenience. Again, it is emphasized that transformation to the rotating frame of operators here would make the coefficients time-dependent in an oscillating manner, and it is far better to be avoided for these classes of nonlinear problems. 

\subsubsection{First-order Perturbation to (\ref{eq16})}

In a similar manner to (\ref{eq13}), we may assume photon and phonon baths each have a mean boson number respectively as $\left<\hat{n}\right>=\bar{n}$ and $\left<\hat{m}\right>=\bar{m}$, which gives
\begin{eqnarray}
\label{eq17}
\dot{\hat{c}}&=&-i4\gamma\bar{m}\left(2\hat{c}+\hat{n}\right)-i4\gamma\left(\bar{n}+\frac{1}{2}\right)(\hat{d}+\hat{d}^\dagger+\hat{m})-\left(\bar{n}+\frac{1}{2}\right)\left(\frac{1}{2}\Gamma_1\hat{c}+\sqrt{\Gamma_1}\hat{c}_{\textrm{in}}\right),\\ \nonumber
\dot{\hat{c}}^\dagger&=&i4\gamma\bar{m}\left(2\hat{c}^\dagger+\hat{n}\right)+i4\gamma\left(\bar{n}+\frac{1}{2}\right)(\hat{d}+\hat{d}^\dagger+\hat{m})-\left(\bar{n}+\frac{1}{2}\right)\left(\frac{1}{2}\Gamma_1\hat{c}^\dagger+\sqrt{\Gamma_1}\hat{c}_{\textrm{in}}^\dagger\right),\\ \nonumber
\dot{\hat{n}}&=&i8\hbar\bar{m}(\hat{c}-\hat{c}^\dagger),\\ \nonumber
\dot{\hat{d}}&=&-i4\gamma\bar{n}\left(2\hat{d}+\hat{m}\right)-i4\gamma\left(\bar{m}+\frac{1}{2}\right)(\hat{c}+\hat{c}^\dagger+\hat{n})-\left(\bar{m}+\frac{1}{2}\right)\left(\frac{1}{2}\Gamma_2\hat{d}+\sqrt{\Gamma_2}\hat{d}_{\textrm{in}}\right),\\ \nonumber
\dot{\hat{d}}^\dagger&=&i4\gamma\bar{n}\left(2\hat{d}^\dagger+\hat{m}\right)+i4\gamma\left(\bar{m}+\frac{1}{2}\right)(\hat{c}+\hat{c}^\dagger+\hat{n})-\left(\bar{m}+\frac{1}{2}\right)\left(\frac{1}{2}\Gamma_2\hat{d}^\dagger+\sqrt{\Gamma_2}\hat{d}_{\textrm{in}}^\dagger\right),\\ \nonumber
\dot{\hat{m}}&=&i8\hbar\gamma\bar{n}(\hat{d}-\hat{d}^\dagger).
\end{eqnarray}

We here need to assume the redefinition $\sqrt{\Delta_1}=(\bar{n}+\frac{1}{2})\sqrt{\Gamma_1}$. Now, without taking $\mathbb{H}_{\rm self}$ into account, this will lead to the linear system of matrix Langevin equations

\begin{flalign}
\nonumber
\begin{bmatrix}
-i8\gamma\bar{m}-\frac{2\bar{n}+1}{4}\Gamma_1 & 0 & -i4\gamma\bar{m} & -i2\gamma(2\bar{n}+1) & -i2\gamma(2\bar{n}+1) & -i2\gamma(2\bar{n}+1) \\
0 & i8\gamma\bar{m}-\frac{2\bar{n}+1}{4}\Gamma_1 & i4\gamma\bar{m} & i2\gamma(2\bar{n}+1) & i2\gamma(2\bar{n}+1) & i2\gamma(2\bar{n}+1) \\
i8\gamma\bar{m} & -i8\gamma\bar{m} & 0 & 0 & 0 & 0 \\ 
-i2\gamma(2\bar{m}+1) & -i2\gamma(2\bar{m}+1) & -i2\gamma(2\bar{m}+1) & -i8\gamma\bar{n}-\frac{2\bar{m}+1}{4}\Gamma_2 & 0 & -4i\gamma\bar{n}\\
i2\gamma(2\bar{m}+1) & i2\gamma(2\bar{m}+1) & i2\gamma(2\bar{m}+1) & 0 & i8\gamma\bar{n}-\frac{2\bar{m}+1}{4}\Gamma_2 & i4\gamma\bar{n}\\
0 & 0 & 0 & i8\gamma\bar{n} & -i8\gamma\bar{n} & 0
\end{bmatrix}
\\ \label{eq18}
\times
\begin{Bmatrix}
\hat{c}\\
\hat{c}^\dagger\\
\hat{n}\\
\hat{d}\\
\hat{d}^\dagger\\
\hat{m}
\end{Bmatrix}
-\begin{Bmatrix}
\sqrt{\Delta_1}\hat{c}_{\textrm{in}}\\
\sqrt{\Delta_1}\hat{c}^\dagger_{\textrm{in}}\\
0\\
\sqrt{\Delta_2}\hat{d}_{\textrm{in}}\\
\sqrt{\Delta_2}\hat{d}^\dagger_{\textrm{in}}\\
0
\end{Bmatrix}=\frac{d}{dt}\begin{Bmatrix}
\hat{c}\\
\hat{c}^\dagger\\
\hat{n}\\
\hat{d}\\
\hat{d}^\dagger\\
\hat{m}
\end{Bmatrix},
\end{flalign}
\noindent
which is, of course, integrable now. The initial cavity boson numbers $\bar{n}$ and $\bar{m}$ can be set in the same manner which was done for the system of equations (\ref{eq14}). Numerical tests reveal that (\ref{eq18}) is conditionally stable if the optical intensity is kept below a certain limit on the red detuning, and is otherwise unstable.  

\subsection{Second order Perturbation to (\ref{eq12},\ref{eq16})}

The set of Langevin equations (\ref{eq12},\ref{eq16}) can be integrated with much more accuracy, if we first identify and sort out the cross terms as individual operators. For instance, (\ref{eq16}) contains the cross operators $\hat{c}\hat{d}$, $\hat{c}\hat{d}^\dagger$, $\hat{c}\hat{m}$, $\hat{c}^\dagger\hat{d}$, $\hat{c}^\dagger\hat{d}^\dagger$, $\hat{c}^\dagger\hat{m}$, $\hat{n}\hat{d}$, $\hat{n}\hat{d}^\dagger$, as well as $\hat{n}\hat{m}$ which is self-adjoint. These constitute an extra set of nine cross operators to be included in the treatment. All these cross operators are formed by multiplication of photonic and phononic single operators, whose notation order, such as $\hat{c}\hat{d}=\hat{d}\hat{c}$ and so on, is obviously immaterial. 

Now, one may proceed first to determine the commutators between these terms where relevant, which always result in linear combinations of the other existing terms. This will clearly enable a more accurate formulation of (\ref{eq16}) but in a $6+9=15$ dimensional space, which is given by the array of operators $\{A\}^{\rm T}=\{\hat{c},\hat{c}^\dagger,\hat{n},\hat{d},\hat{d}^\dagger,\hat{m},\hat{c}\hat{d},\hat{c}\hat{d}^\dagger,\hat{c}\hat{m},\hat{c}^\dagger\hat{d},\hat{c}^\dagger\hat{d}^\dagger,\hat{c}^\dagger\hat{m},\hat{n}\hat{d},\hat{n}\hat{d}^\dagger,\hat{n}\hat{m}\}$. 

The independent non-trivial quadratic commutator equations among cross operators here are found after tedious but straightforward algebra as
\begin{eqnarray}
\nonumber
[\hat{c}\hat{d},\hat{c}^\dagger\hat{d}^\dagger]&=&\frac{1}{8}[(2\hat{n}\hat{m}+3)(\hat{m}+\hat{n}+2)+\hat{n}^2+\hat{m}^2-4],\\ \nonumber
[\hat{c}\hat{d},\hat{c}^\dagger\hat{m}]&=&\frac{1}{2}(\hat{n}^2+2\hat{n}\hat{m}+2\hat{n}+\hat{m}+2)\hat{d},\\ \nonumber
[\hat{c}\hat{d},\hat{n}\hat{d}^\dagger]&=&\frac{1}{2}(\hat{m}^2+3\hat{m}+2\hat{m}\hat{n}+\hat{n}+2)\hat{c},\\ \nonumber
[\hat{c}\hat{d},\hat{n}\hat{m}]&=&(\hat{n}+\hat{m}+4)\hat{c}\hat{d},\\ \nonumber
[\hat{c}\hat{d}^\dagger,\hat{c}^\dagger\hat{d}]&=&\frac{1}{8}(2\hat{n}\hat{m}+\hat{m}+\hat{n}-1)(\hat{m}-\hat{n}),\\ \nonumber
[\hat{c}\hat{d}^\dagger,\hat{c}^\dagger\hat{m}]&=&\frac{1}{2}\left[(2\hat{n}+1)\hat{m}-(\hat{n}+1)(\hat{n}+2)\right]\hat{d}^\dagger,\\ \nonumber
[\hat{c}\hat{d}^\dagger,\hat{n}\hat{d}]&=&\frac{1}{2}\left[\hat{m}(\hat{m}-2\hat{n})-(\hat{m}+\hat{n})\right]\hat{c},\\ \nonumber
[\hat{c}\hat{d}^\dagger,\hat{n}\hat{m}]&=&2(\hat{m}-\hat{n}-2)\hat{c}\hat{d}^\dagger,\\ \nonumber
[\hat{c}\hat{m},\hat{n}\hat{d}]&=&2(\hat{m}+\hat{n}+2)\hat{c}\hat{d},\\ \label{eq19}
[\hat{c}\hat{m},\hat{n}\hat{d}^\dagger]&=&2(\hat{m}+\hat{n})\hat{c}\hat{d}^\dagger.
\end{eqnarray}
\noindent
The rest of commutators among cross operators are either adjoints of the above, or have a common term which makes their evaluation possible using either (\ref{eq4}) or (\ref{eq5}). Commutators among cross operators and single operators can be always factored, such as $[\hat{c}\hat{d},\hat{n}]=[\hat{c},\hat{n}]\hat{d}$. Commutators among single operators are already known (\ref{eq4},\ref{eq5}). It can be therefore seen that commutators (\ref{eq19}) always lead to operators of higher orders yet, so that they do not terminate at any finite order of interest by merely expansion of operators basis. This fact puts the perturbative method put into work. There are, however, nonlinear systems such as semiconductor optical cavities \cite{33.19,33b} in which higher-order operators yield an exact closed algebra and satisfy a closedness property within the original space by appropriate definition. 

The set of ten commutators now can be perturbatively linearized as a second-order approximation, by replacing the number operators with their mean values, wherever needed to reduce the set of operators back to the available 15 dimensional space. This will give rise to the similar set of equations after some algebra 
\begin{eqnarray}
\nonumber
[\hat{c}\hat{d},\hat{c}^\dagger\hat{d}^\dagger]&=&\frac{1}{16}(\bar{m}+\bar{n}+8)\hat{n}\hat{m}+\frac{1}{8}\left[\bar{m}(\bar{n}+1)+\frac{1}{2}\bar{n}^2+3\right]\hat{m}+\frac{1}{8}\left[\bar{n}(\bar{m}+1)+\frac{1}{2}\bar{m}^2+3\right]\hat{n}+\frac{1}{4},\\ \nonumber
[\hat{c}\hat{d},\hat{c}^\dagger\hat{m}]&=&\frac{1}{2}(\bar{n}+2\bar{m}+2)\hat{n}\hat{d}+\frac{1}{2}(\bar{m}+2)\hat{d},\\ \nonumber
[\hat{c}\hat{d},\hat{n}\hat{d}^\dagger]&=&\frac{1}{2}(\bar{m}+3+2\bar{n})\hat{c}\hat{m}+\frac{1}{2}(\bar{n}+2)\hat{c},\\ \nonumber
[\hat{c}\hat{d},\hat{n}\hat{m}]&=&(\bar{n}+\bar{m}+4)\hat{c}\hat{d},\\ \nonumber
[\hat{c}\hat{d}^\dagger,\hat{c}^\dagger\hat{d}]&=&\frac{1}{16}(\bar{m}-\bar{n})\hat{n}\hat{m}+\frac{1}{8}\left[\bar{m}(\bar{n}+1)-1-\frac{1}{2}\bar{n}^2\right]\hat{m}-\frac{1}{8}\left[\bar{n}(\bar{m}+1)-1-\frac{1}{2}\bar{m}^2\right]\hat{n},\\ \nonumber
[\hat{c}\hat{d}^\dagger,\hat{c}^\dagger\hat{m}]&=&\frac{1}{2}(2\bar{m}-\bar{n}-3)\hat{n}\hat{d}^\dagger+\frac{1}{2}(\bar{m}-2)\hat{d}^\dagger,\\ \nonumber
[\hat{c}\hat{d}^\dagger,\hat{n}\hat{d}]&=&\frac{1}{2}(\bar{m}-2\bar{n}-1)\hat{c}\hat{m}-\frac{1}{2}\bar{n}\hat{c},\\ \nonumber
[\hat{c}\hat{d}^\dagger,\hat{n}\hat{m}]&=&2(\bar{m}-\bar{n}-2)\hat{c}\hat{d}^\dagger,\\ \nonumber
[\hat{c}\hat{m},\hat{n}\hat{d}]&=&2(\bar{m}+\bar{n}+2)\hat{c}\hat{d},\\ \label{eq20}
[\hat{c}\hat{m},\hat{n}\hat{d}^\dagger]&=&2(\bar{m}+\bar{n})\hat{c}\hat{d}^\dagger.
\end{eqnarray}
\noindent
where the reduction of triple operator products among single and cross operators as  $4\hat{x}\hat{y}\hat{z}\rightarrow\bar{x}\hat{y}\hat{z}+\bar{x}\bar{y}\hat{z}+\bar{y}\bar{z}\hat{x}+\bar{z}\bar{x}\hat{y}$ is used where appropriate. For instance, the term $4\hat{n}\hat{m}^2$ is replaced as $\bar{m}\hat{m}\hat{n}+2\bar{m}\bar{n}\hat{m}+\bar{m}^2\hat{n}$ and so on. Also, similar to (\ref{eq13}), products among single operators are reduced as $2\hat{x}\hat{y}\rightarrow\bar{x}\hat{y}+\bar{y}\hat{x}$. This is somewhat comparable to the mean field approach in cross Kerr optomechanics \cite{38a}.

There are two basic reasons why we have adopted this particular approach to the linearization and cuting off the diverging operators of higher orders. The first reason is that number operators vary slowly in time as opposed to their bosonic counterparts which oscillate rapidly in time, given the fact that the use of rotating frames is disallowed here. Secondly, number operators are both positive-definite and self-adjoint, and thus can be approximated by a positive real number. These properties makes the replacements $\hat{n}\rightarrow\bar{n}$ and $\hat{m}\rightarrow\bar{m}$ reasonable approximations, and the replacement with mean values needs only to be restricted to the number operators, to yield a closed algebra necessary for construction of Langevin equations. Hence, the correct application of replacements only to the triple operator products appearing in (\ref{eq19}) will make sure that no operator having an order beyond than that of cross operators will appear in the formulation.

Anyhow, it can be seen now that all approximate commutators in (\ref{eq20}) allow the set of operators $\{A\}^{\rm T}\cup\{\hat{1}\}=\{\hat{1},\hat{c},\hat{c}^\dagger,\hat{n},\hat{d},\hat{d}^\dagger,\hat{m},\hat{c}\hat{d},\hat{c}\hat{d}^\dagger,\hat{c}\hat{m},\hat{c}^\dagger\hat{d},\hat{c}^\dagger\hat{d}^\dagger,\hat{c}^\dagger\hat{m},\hat{n}\hat{d},\hat{n}\hat{d}^\dagger,\hat{n}\hat{m}\}$ to take on linear combinations of its members among every pair of commutations possible, where $\hat{1}$ is the identity operator. Obviously, this approximate closedness property now makes the full construction of Langevin equations for the operators belonging to $\{A\}$ possible. It is noted that $\hat{1}$ is not an identity element for the commutation.

We can now define the set $\{S\}={\rm span}(\{A\}\cup\{\hat{1}\})$, which is spanned by all possible linear combinations of $\{\hat{1}\}$ and the members of $\{A\}$ together with the associative binary commutation operation $[]$ defined in (\ref{eq4},\ref{eq5},\ref{eq20}). The ordered pair $(\{{S}\},[])$ is now a semigroup.

Having therefore these ten commutators (\ref{eq20}) known, we may proceed now to composing the second-order approximation to the nonlinear Langevin equations (\ref{eq16}), from which a much more accurate solution could be obtained. Here, the corresponding Langevin equations may be constructed at each step by setting both $\hat{z}$ and $\hat{x}$ in (\ref{eq10}) equal to either of the 15 operators, while the noise input terms for cross operators is a simple product of related individual noise terms. The linear damping rates of higher-order operators is furthermore simply the sum of individual damping rates of corresponding single operators, which completes the needed parameter set of Langevin equations.

\section{Further Considerations}
\subsection{Optomechanical Interaction \& Drive Terms}

The method described in the above can be simultaneously used if other terms such as the standard optomechanical interaction $\mathbb{H}_{\textrm{OM}}$ is non-zero, or there exists a coherent pumping drive term which can be expressed as  $\mathbb{H}_{\textrm{d}}=\sum_k F_k \hat{b}^\dagger+F_k^*\hat{b}$, where $F_k$ are time-dependent drive amplitudes. While $\mathbb{H}_{\textrm{d}}$ does not appear directly in the Langevin equations, treatment of $\mathbb{H}_{\textrm{OM}}$ requires inclusion of additional Langevin equations for $\hat{a}$ and $\hat{b}$ where appropriate, as well as few extra terms in the rest. This can be done in a pretty standard way, and is not repeated here for the sake of brevity \cite{1,2,24,25,26,26a}. 

\subsection{Multi-mode Fields}

The analysis is also essentially unaltered if there are more than one mechanical mode to be considered \cite{17,39,39a}, and the method is still easily applicable with no fundamental change. Suppose that there are a total of $M$ mechanical modes with the corresponding bosonic operators $\hat{b}_k$ and $\hat{b}_k^\dagger$ where $k\in[1,M]$. Then, these modes are mutually independent in the sense that $[\hat{b}_j,\hat{b}_k]=0$ and $[\hat{b}_j,\hat{b}_k^\dagger]=\delta_{jk}$. The set of commutators (\ref{eq5}) will be usable for all $M$ modes individually and as a result (\ref{eq19}) and therefore (\ref{eq20}) may be still used. The first and second order perturbations will respectively result in $3+3M=3(M+1)$ and $3+3M+9M=3(4M+1)$ equations. The redefined set of operators will be respectively now $\{A\}^{\rm T}=\{\hat{c},\hat{c}^\dagger,\hat{n},\hat{d}_k,\hat{d}_k^\dagger,\hat{m}_k;k\in[1,M]\}$ and $\{A\}^{\rm T}=\{\hat{c},\hat{c}^\dagger,\hat{n},\hat{d}_k,\hat{d}_k^\dagger,\hat{m}_k,\hat{c}\hat{d}_k,\hat{c}\hat{d}_k^\dagger,\hat{c}\hat{m}_k,\hat{c}^\dagger\hat{d}_k,\hat{c}^\dagger\hat{d}_k^\dagger,\hat{c}^\dagger\hat{m}_k,\hat{n}\hat{d}_k,\hat{n}\hat{d}_k^\dagger,\hat{n}\hat{m}_k;k\in[1,M]\}$.

Similarly, in case of $N$ optical modes satisfying $[\hat{a}_j,\hat{a}_k]=0$ and $[\hat{a}_j,\hat{a}_k^\dagger]=\delta_{jk}$, the set of commutators (\ref{eq4}) can be used and the operator set should be now expanded as $\{A\}^{\rm T}=\{\hat{c}_j,\hat{c}_j^\dagger,\hat{n}_j,\hat{d}_k,\hat{d}_k^\dagger,\hat{m}_k;j\in[1,N];k\in[1,M]\}$ and $\{A\}^{\rm T}=\{\hat{c}_j,\hat{c}_j^\dagger,\hat{n}_j,\hat{d}_k,\hat{d}_k^\dagger,\hat{m}_k,\hat{c}_j\hat{d}_k,\hat{c}_j\hat{d}_k^\dagger,\hat{c}_j\hat{m}_k,\hat{c}_j^\dagger\hat{d}_k,\hat{c}_j^\dagger\hat{d}_k^\dagger,\hat{c}_j^\dagger\hat{m}_k,\hat{n}_j\hat{d}_k,\hat{n}_j\hat{d}_k^\dagger,\hat{n}_j\hat{m}_k;j\in[1,N];k\in[1,M]\}$ respectively for first and second order perturbations. Hence, the corresponding dimensions will be now respectively either $3(N+M)$ or $3(N+M+3NM)$. Higher-order commutators (\ref{eq19}) and (\ref{eq20}) can be still used again by only addition of appropriate photonic $j$ and phononic $k$ mode indices to the respective operators contained in the expanded operator basis set $\{A\}$.

\subsection{Noise Spectra}\label{NoiseSpectra}

The required noise spectra \cite{39b} of cross operators is clearly a product of each of the individual terms, since the nature of particles are different. However, the noise spectra of quadratic operators themselves need to be appropriately expressed. For instance, $\hat{d}_{\textrm{in}}$ actually corresponds to the spectral input noise of the square operator $\hat{d}=\hat{b}\hat{b}/2\sqrt{\Gamma}$ from (\ref{eq3}), which clearly satisfies $\hat{d}_{\textrm{in}}(t)=\frac{1}{2}\hat{b}_{\textrm{in}}(t)\hat{b}_{\textrm{in}}(t)/\sqrt{\Gamma}$, or $\hat{d}_{\textrm{in}}(w)=\frac{1}{2}\hat{b}_{\textrm{in}}(w)\ast \hat{b}_{\textrm{in}}(w)/\sqrt{\Gamma}$ in the frequency domain, where $\ast$ merely represents the convolution operation. Therefore, once $\hat{a}_{\textrm{in}}(w)$ and $\hat{b}_{\textrm{in}}(w)$ are known, all relevant remaining input noise spectra could be obtained accordingly using simple convolutions or products in frequency domain.

As a result, the corresponding spectral density of the noise input terms to the cross operators can be determined from the relevant vacuum noise fluctuations and performing a Fourier transform. For instance, we have $S_{CDCD}[w]=S_{CC}[w]S_{DD}[w]$ where $S_{CC}[w]=\frac{1}{4}S_{A^2A^2}[w]$ and $S_{DD}[w]=\frac{1}{4}S_{B^2B^2}[w]$. Then Isserlis-Wick theorem \cite{32,40} could be exploited to yield the desired expressions. If we assume 
\begin{eqnarray}
\label{eq21}
\left<\hat{f}(t)\hat{f}(\tau)\right>&=&\zeta(t-\tau), \\ \nonumber
\left<\hat{f}(t)\hat{f}^\dagger(\tau)\right>&=&\psi(t-\tau), \\ \nonumber
[\hat{f}(t),\hat{f}^\dagger(\tau)]&=&\hat{\upsilon}(t-\tau),
\end{eqnarray}  where the dimensionless correlation integrator runs on phase, instead of time, as
\begin{equation} 
\label{eq22}
\left<\hat{f}(t)\hat{g}(\tau)\right>=\int \hat{f}(t+\tau)\hat{g}(\tau) d(\omega \tau),
\end{equation}
\noindent
then the functions $\zeta(\cdot),$ $\psi(\cdot),$ and the operator $\hat{v}(\cdot)$ should be all having the dimension of $\hat{f}^2(\cdot)$ as well. That means if $\hat{f}$ is dimensionless, which is the case for the choice of ladder operators, then $\zeta(\cdot),$ $\psi(\cdot),$ and $\hat{v}(\cdot)$ become dimensionless, too. The functions $\zeta(\cdot)$ and $\psi(\cdot)$ together can cause squeezing or thermal states if appropriately defined \cite{24,26a}. By Isserlis-Wick theorem applied to scalars we have $\left<x_1 x_2 x_3 x_4\right>=\left<x_1 x_2\right>\left<x_3 x_4\right>+\left<x_1 x_3\right>\left<x_2 x_4\right>+\left<x_1 x_4\right>\left<x_3 x_4\right>$. This gives for the operators
\begin{eqnarray}
\label{eq23}
S_{F^2F^2}[w]&=&\frac{1}{2\pi}\int_{-\infty}^{+\infty}\left<\hat{f}^2(t)\hat{f}^{2\dagger}(0)\right>e^{iwt} dt =\frac{1}{2\pi}\int_{-\infty}^{+\infty}\left<\hat{f}(t)\hat{f}(t)\hat{f}^{\dagger}(0)\hat{f}^{\dagger}(0)\right>e^{iwt} dt \\ \nonumber
&=&\frac{1}{2\pi}\int_{-\infty}^{\infty}\left\{\left<\hat{f}^2(t)\right>\left<\hat{f}^{2\dagger}(0)\right>+2\left<\hat{f}(t)\hat{f}^\dagger(0)\right>^2+2\left<\hat{f}(t)\left[\hat{f}(t),\hat{f}^\dagger(0)\right]\hat{f}^\dagger(0)\right>\right\} e^{iwt} dt \\ \nonumber
&=&\frac{1}{2\pi}\int_{-\infty}^{\infty}\left[\zeta(0)\zeta^\ast(0)+2\psi^2(t)+2\left<\hat{f}(t)\hat{\upsilon}(t)\hat{f}^\dagger(0)\right>\right] e^{iwt} dt\\ \nonumber 
&=&|\zeta(0)|^2\delta(w)+\frac{1}{\pi}\int_{-\infty}^{\infty}\left[\psi^2(t)+\left<\hat{f}(t)\hat{\upsilon}(t)\hat{f}^\dagger(0)\right>\right] e^{iwt} dt. 
\end{eqnarray} 
\noindent
Hence, for a given stochastic process where $\left<\hat{f}(t)\hat{f}(\tau)\right>=0$, $\left<\hat{f}(t)\hat{f}^\dagger(\tau)\right>=\Psi(t-\tau)$, and having the scalar commutator $[\hat{f}(t),\hat{f}^\dagger(\tau)]=\Upsilon (t-\tau)$, we simply get
\begin{equation}
\label{eq24}
S_{F^2F^2}[w]=\frac{1}{\pi}\int_{-\infty}^{\infty}\Psi^2(t) e^{iwt} dt+\frac{1}{\pi}\int_{-\infty}^{\infty}\Upsilon (t)\Psi(t) e^{iwt} dt. 
\end{equation} 

Now, suppose that we have a coherent field of photons at the angular frequency $\omega$ with an initial Gaussian distribution, in which $\Psi(t)=\exp(-\chi^2\omega^2 t^2/2)\exp(-i\omega t)$ and $\Upsilon (t)=\Psi(t)$, while having the linewidth $\Delta f=\frac{1}{2\pi}\chi\omega$. Clearly, $\chi$ is a dimensionless and positive real number. In the limit of $\chi\rightarrow 0^{+}$, the expected relationship $\Psi(t)=\sqrt{2\pi}\delta(\omega t)/\chi$ is easily recovered. 

This particular definition of the correlating function $\Psi(t)$ ensures that the corresponding spectral density is appropriately normalized, that is
\begin{eqnarray}
\label{eq25}
\int_{-\infty}^{+\infty} S_{FF}[w]dw&=&\int_{-\infty}^{\infty}\left[\frac{1}{2\pi}\int_{-\infty}^{+\infty}\left<\hat{f}(t)\hat{f}^\dagger(0)\right>e^{iwt} dt\right]dw\\ \nonumber
&=&1.
\end{eqnarray}
\noindent
Hence, one may obtain the following spectral density
\begin{equation}
\label{eq26}
S_{F^2F^2}[w]=\frac{\chi}{\pi\sqrt{\pi}\omega}\exp\left[-\frac{(w-2\omega)^2}{4\chi^2\omega^2}\right],
\end{equation}
\noindent
which is centered at the doubled frequency $2\omega$, has a linewidth of $\sqrt{2}\Delta f$, and satisfies the property 
\begin{equation}
\label{eq27}
\int_{-\infty}^{+\infty} S_{F^2F^2}[w]dw=\frac{2}{\pi}\chi^2.
\end{equation}

Once the spectral densities of input noise terms are found, spectral densities of all output fields immediately follows (\ref{eq8},\ref{eq9}) as $\{A[w]\}_{\rm out}=[\textbf{S}^\dagger(w)\textbf{S}(w)]\{A[w]\}_{\rm in}$, in which $[\textbf{S}^\dagger(w)\textbf{S}(w)]=[|S_{ij}(w)|^2]$,  $\{A[w]\}_{\rm in}$ is an array containing the spectral densities of inputs, and similarly $\{A[w]\}_{\rm out}$ is the array of spectral densities at each of the output fields.

\subsection{Estimation of $g^{(2)}(0)$}

Many of the important features of an interacting quantum system is given by its second-order correlation function $g^{(2)}(0)$ at zero time-delay \cite{36a,36b,36c} defined as
\begin{equation}
\label{eq28}
g^{(2)}(0)=\frac{\langle\hat{a}^\dagger(0)\hat{a}^\dagger(0)\hat{a}(0)\hat{a}(0)\rangle}{\langle \hat{a}^\dagger(0)\hat{a}(0)\rangle^2}.
\end{equation}
\noindent
It is fairly easy to estimate this function once the spectral densities of all higher order operators of the nonlinear system are calculated. For this purpose, we may first employ the definition (\ref{eq3}) to rewrite
\begin{equation}
\label{eq29}
g^{(2)}(0)=4\frac{\langle\hat{c}^\dagger(0)\hat{c}(0)\rangle}{\langle \hat{n}(0)\rangle^2}=\frac{4}{\bar{n}^2}\langle\hat{c}^\dagger(0)\hat{c}(0)\rangle=\frac{4}{\bar{n}^2}\left[\langle\hat{c}(0)\hat{c}^\dagger(0)\rangle-\bar{n}-\frac{1}{2}\right].
\end{equation}
\noindent
Estimation of the average within brackets can be done by having $S_{CC}[w]=\frac{1}{4}S_{A^2A^2}[w]$ corresponding to the higher-order operator $\hat{c}$. This can be assumed to has been already found from knowledge of the scattering matrix $[\textbf{S}(w)]$, spectral densities of input fields $\{A[w]\}_{\rm in}$, and subsequent derivation of spectral density array of output fields $\{A[w]\}_{\rm out}$. Then, $S_{CC}[w]$ will be simply an element of the vector $\{A[w]\}_{\rm out}$. Using (\ref{eq24}), this results in a fairly brief representation 
\begin{equation}
\label{eq30}
g^{(2)}(0)=\frac{4}{\bar{n}^2}\left(\int_{-\infty}^{+\infty}S_{CC}[w]dw\right)-\frac{4\bar{n}+2}{\bar{n}^2}=\frac{2}{\bar{n}^2}\Psi(0)\left[\Psi(0)+\Upsilon(0)\right]-\frac{4\bar{n}+2}{\bar{n}^2}.
\end{equation}
\noindent
With the assumptions above for an ideal initial Gaussian distribution, we have $\Psi(0)=\Upsilon(0)=1$ and thus $g^{(2)}(0)=4(\frac{1}{2}-\bar{n})/\bar{n}^2$. One should have in mind that this relationship cannot be readily used for a coherent radiation, since for a practical laser the true statistics is Poissonian and not Gaussian. This analysis thus reveals that the cavity occupation number of such an ideal laser with the threshold defined as $g^{(2)}(0)=1$ is exactly $\bar{n}=\sqrt{6}-2\approx 0.450$. This is in contrast to the widely used assumption of quantum threshold condition $\bar{n}=1$ \cite{40a,40b,40b1,40b2,40b3,40b4}. Interestingly, a new study \cite{40c} of photon statistics in weakly nonlinear optical cavities based on extensive density matrix calculations \cite{40d,40e} yields the value $\bar{n}=0.4172$, which is in reasonable agreement to our estimate. An earlier investigation on quantum-dot photonic crystal cavity lasers \cite{40f,40g} also gives the value $\bar{n}=0.485$.

\section{Anharmonic Oscillator}

The quantum anharmonic oscillator appears in many nonlinear systems including quadratic optomechanics \cite{40h,40i}, where our method here is applicable. The anharmonic Kerr Hamiltonian is \cite{41,42}
\begin{equation}
\label{eq31}
\mathbb{H}=\hbar\omega\hat{a}^\dagger\hat{a}+\frac{1}{2}\hbar\zeta\hat{a}^{\dagger 2}\hat{a}^2=\hbar\omega\hat{a}^\dagger\hat{a}+2\hbar\zeta\hat{c}^{\dagger}\hat{c}=\hbar\left(\omega-\frac{1}{2}\zeta\right)\hat{n}+\frac{1}{2}\hbar\zeta\hat{n}^2,
\end{equation}
\noindent
in which $\zeta$ is a constant. It is well known that in case of $\zeta>2\omega$ this system exhibits an effective bistable potential, and is otherwise monostable. However, we are here much interested in a slightly different but more complicated form given by \cite{2DM}
\begin{equation}
\label{eq32}
\mathbb{H}=\hbar\omega\hat{a}^\dagger\hat{a}-\frac{1}{2}\hbar\zeta\left(\hat{a}^\dagger+\hat{a}\right)^4,
\end{equation}
\noindent
which is monostable or bistable if both $\omega$ and $\zeta$ are respectively positive or negative. This type of nonlinearity is of particular importance in fourth-order analysis of qubits \cite{43,44,45,46,47,49,50}. While the Hamiltonian (\ref{eq32}) is for a single-mode field, the case of multi-mode electromagnetic field could be easily devised following the existing interaction Hamiltonians \cite{2DM} and the presented method in this article. Nevertheless, the above expression after some algebraic manipulations can be put into the form 
\begin{equation}
\label{eq33}
\mathbb{H}=\hbar(\omega-3\zeta)\hat{n}-3\hbar\zeta\hat{n}^2-2\hbar\zeta\left[\hat{c}^2+\hat{c}^{\dagger 2}+3 \left(\hat{c}+\hat{c}^\dagger\right)\right]-4\hbar\zeta\left(\hat{n}\hat{c}+\hat{c}^\dagger\hat{n}\right),
\end{equation}
\noindent
where a trivial constant term $\hbar\zeta$ is dropped. Here, we may proceed with the 8-dimensional basis operator set $\{A\}^{\rm T}=\{\hat{c},\hat{c}^\dagger,\hat{n},\hat{n}^2,\hat{c}^2,\hat{c}^{\dagger 2},\hat{n}\hat{c},\hat{c}^\dagger\hat{n}\}$, resulting in a second order perturbation accuracy.

Treating this problem using the Langevin equation (\ref{eq10}), regardless of the values of $\zeta$ and $\omega$, is possible, only if the following non-trivial exact commutators
\begin{eqnarray}
\label{eq34}
[\hat{n},\hat{c}^2]&=&-4\hat{c}^2, \\ \nonumber
[\hat{n}^2,\hat{c}]&=&-3\hat{n}\hat{c}-\frac{7}{2}\hat{c}, \\ \nonumber
[\hat{n}^2,\hat{c}^2]&=&4(\hat{n}-2)\hat{n}\hat{c}^2, \\ \nonumber
[\hat{c}^2,\hat{c}^\dagger]&=&2\hat{n}\hat{c}+3\hat{c}, \\ \nonumber
[\hat{c}^2,\hat{c}^{\dagger 2}]&=&\hat{n}^3+\frac{3}{2}\left(\hat{n}^2+1\right)+\frac{1}{4}\hat{n}, \\ \nonumber
[\hat{c}^2,\hat{c}^\dagger\hat{n}]&=& 3\left(\hat{n}+2\right)\hat{n}\hat{c}+6\hat{c},\\ \nonumber
[\hat{c},\hat{c}^\dagger\hat{n}]&=&\frac{3}{2}\hat{n}^2, \\ \nonumber
[\hat{n}\hat{c},\hat{c}^\dagger\hat{n}]&=&\frac{1}{2}\left(4\hat{n}^2-3\hat{n}+2\right)\hat{n},
\end{eqnarray}
\noindent
are known, which may be found after significant algebra. The rest of required commutators which are not conjugates of those in the above, can either directly or after factorization of a common term be easily found from (\ref{eq4}). Again, the set of commutators (\ref{eq34}) does not yet satisfy the closedness property within $\{S\}={\rm span}(\{A\}\cup\{\hat{1}\})$, unless the approximate linearization 

\begin{eqnarray}
\label{eq35}
[\hat{n},\hat{c}^2]&=&-4\hat{c}^2, \\ \nonumber
[\hat{n}^2,\hat{c}]&=&-3\hat{n}\hat{c}-\frac{7}{2}\hat{c}, \\ \nonumber
[\hat{n}^2,\hat{c}^2]&=&4(\bar{n}-2)\bar{n}\hat{c}^2, \\ \nonumber
[\hat{c}^2,\hat{c}^\dagger]&=&2\hat{n}\hat{c}+3\hat{c}, \\ \nonumber
[\hat{c}^2,\hat{c}^{\dagger 2}]&=&\frac{1}{2}\left(2\bar{n}+3\right)\hat{n}^2+\frac{1}{4}\hat{n}+\frac{3}{2}, \\ \nonumber
[\hat{c}^2,\hat{c}^\dagger\hat{n}]&=& 3\left(\bar{n}+2\right)\hat{n}\hat{c}+6\hat{c},\\ \nonumber
[\hat{c},\hat{c}^\dagger\hat{n}]&=&\frac{3}{2}\hat{n}^2, \\ \nonumber
[\hat{n}\hat{c},\hat{c}^\dagger\hat{n}]&=&\frac{1}{2}\left(4\bar{n}-3\right)\hat{n}^2+\hat{n},
\end{eqnarray}
\noindent
is employed. The rest of the process is identical to the one described under (\ref{eq20}). Construction of the respective noise terms is also possible by iterated use of the results in \S \ref{NoiseSpectra} and so on.

\subsection{The Husimi-Kano Q-functions}

It is mostly appropriate that moments of operators are known, which are scalar functions and much easier to work with. The particular choice of $Q-$functions \cite{50a} is preferred when dealing with ladder operators, and are obtained by taking the expectation value of density operator with respect to a complex coherent state $\ket{\alpha}$ and dividing by $\pi$. This definition leads to a non-negative real valued function $Q(\alpha)=Q(\Re[\alpha],\Im[\alpha])$ of $\ket{\alpha}$. Then, obtaining $Q-$function moments of any expression containing the ladder operators would be straightforward \cite{50a}. However, it must be antinormally ordered, with creators be moved to the right. In $\{A\}^{\rm T}$ above all operators are actually in the normal form, except $\hat{n}^2$. It is possible to put the nontrivial members of $\{A\}$ in the antinormal order
\begin{eqnarray}
\label{eq35a}
\hat{n}&=&\hat{a}\hat{a}^\dagger-1,\\ \nonumber
\hat{n}^2&=&\hat{a}\hat{a}\hat{a}^\dagger\hat{a}^\dagger-2\hat{a}\hat{a}^\dagger,\\ \nonumber
\hat{n}\hat{c}&=&\frac{1}{2}\hat{a}\hat{a}\hat{a}\hat{a}^\dagger-\frac{3}{2}\hat{a}\hat{a},\\ \nonumber
\hat{c}^\dagger\hat{n}&=&\frac{1}{2}\hat{a}\hat{a}^\dagger\hat{a}^\dagger\hat{a}^\dagger-\frac{3}{2}\hat{a}^\dagger\hat{a}^\dagger.
\end{eqnarray}
While evaluating $Q-$function moments, $\hat{a}$ and $\hat{a}^\dagger$ are replaced with $\alpha$ and $\alpha^\ast$ respectively as
\begin{eqnarray}
\label{eq35b}
\braket{\hat{n}}&=&|\alpha|^2-1,\\ \nonumber
\braket{\hat{n}^2}&=&|\alpha|^4-2|\alpha|^2,\\ \nonumber
\braket{\hat{n}\hat{c}}&=&\frac{1}{2}\alpha^2|\alpha|^2-\frac{3}{2}\alpha^2,\\ \nonumber
\braket{\hat{c}^\dagger\hat{n}}&=&\frac{1}{2}\alpha^{\ast 2}|\alpha|^2-\frac{3}{2}\alpha^{\ast 2}.
\end{eqnarray}
\noindent
All remains now is to redefine the array of $Q-$functions bases, using common terms as $\{\braket{A}\}^{\rm T}=\{\alpha^2,\alpha^{\ast 2},|\alpha|^2,|\alpha|^4,\alpha^4$, $\alpha^{\ast 4},\alpha^2|\alpha|^2,\alpha^{\ast 2}|\alpha|^2\}$ from which the original $Q-$functions could be readily restored. This translates into a set of scalar differential equations which conveniently could be solved. Fluctuations of noise terms also vanish while taking the expectation values, and only their average values survive. To illustrate this, suppose that the system is driven by a coherent field $\hat{a}_{\rm in}$ with the normalized electric field amplitude $\beta=\alpha/\sqrt{2}$ and at the frequency $\omega$. Then, the $Q-$function moments of the input fields after defining the loss rates $\Gamma_3=2\Gamma_2=4\Gamma_1$ become $\braket{\hat{a}_{\rm in}}=\sqrt{2\Gamma_1}\beta$, $\braket{\hat{c}_{\rm in}}=\sqrt{\Gamma_2}\beta$, $\braket{\hat{n}_{\rm in}}=\sqrt{\Gamma_2}(2|\beta|^2+1)$, $\braket{\hat{c}^2_{\rm in}}=\sqrt{\Gamma_3}\beta^2$, and $\braket{\hat{n}_{\rm in}\hat{c}_{\rm in}}=\sqrt{\Gamma_3}\beta^2(2|\beta|^2+3)$.

\subsection{Quantum Limited Amplifiers}
The same method can be extended to the quantum limited amplifiers, which in the general form coincides with the expression (\ref{eq31}), but is usually solved using a zeroth-order perturbation \cite{36e}. For the single-mode degenerate quantum limited amplifier \cite{5,36e,36e1}, the corresponding Hamiltonian is slightly different given by $\mathbb{H}=\hbar\omega\hat{n}+\hbar(g\hat{c}+g^\ast\hat{c}^\dagger)$, with the 3-dimensional basis $\{A\}^{\rm T}=\{\hat{n},\hat{c},\hat{c}^\dagger\}$ which satisfies closedness. Then, the second-order accurate Langevin equations with inclusion of the self-energy $\mathbb{H}_{\rm self}=\hbar\omega\hat{n}$ can be shown to be unconditionally stable with $\Re\{{\rm eig}[{\bf M}]\}<0$, given by
\begin{eqnarray}
\dot{\hat{n}}&=&-i2(g\hat{c}-g^\ast\hat{c}^\dagger),\\ \nonumber
\dot{\hat{c}}&=&(-2i\omega-\frac{2\bar{n}+1}{4}\Gamma_2)\hat{c}-ig^\ast\hat{n}-i\frac{1}{2}g^\ast-(\bar{n}+\frac{1}{2})\sqrt{\Gamma_2}\hat{c}_{\rm in},\\ \nonumber
\dot{\hat{c}}^\dagger&=&(2i\omega-\frac{2\bar{n}+1}{4}\Gamma_2)\hat{c}^\dagger+ig\hat{n}+i\frac{1}{2}g-(\bar{n}+\frac{1}{2})\sqrt{\Gamma_2}\hat{c}^\dagger_{\rm in}.
\end{eqnarray}
In presence of Kerr nonlinearity \cite{36f} as $\mathbb{H}=\hbar\omega\hat{n}+\hbar(g\hat{c}+g^\ast\hat{c}^\dagger)+\hbar\gamma\hat{c}^\dagger\hat{c}$, one may use $4\hat{c}^\dagger\hat{c}=\hat{n}^2-\hat{n}$, $[\hat{n}^2,\hat{c}]\approx-\frac{1}{2}(6\bar{n}+7)\hat{c}$, and the basis $\{A\}^{\rm T}=\{\hat{n},\hat{n}^2,\hat{c},\hat{c}^\dagger\}$ to construct a set of $4\times 4$ integrable Langevin equations. The rest of necessary commutators are already found in (\ref{eq4}), (\ref{eq34}), and (\ref{eq35}).

\subsection{Quantum Nondemolition Measurements}
Quantum nondemolition measurements of states require a cross-Kerr nonlinear interaction of the type $\mathbb{H}=\hbar\omega\hat{a}^\dagger\hat{a}+\hbar\Omega\hat{b}^\dagger\hat{b}+\hbar\chi\hat{a}^\dagger\hat{a}\hat{b}^\dagger\hat{b}=\hbar\omega\hat{n}+\hbar\Omega\hat{m}+\hbar\chi\hat{n}\hat{m}$, in which $\hat{a}$ and $\hat{b}$ fields respectively correspond to the probe and signal \cite{36g,36h}. This system can be conveniently analyzed by the preferred choice \cite{36g} of the higher-order operators $\{A\}^{\rm T}=\{\hat{n},\hat{m},\hat{C},\hat{S}\}$, where
\begin{eqnarray}
\hat{C}&=&\frac{1}{2}\left[(\hat{n}+1)^{-\frac{1}{2}}\hat{a}+\hat{a}^\dagger(\hat{n}+1)^{-\frac{1}{2}}\right], \\ \nonumber
\hat{S}&=&\frac{1}{2i}\left[(\hat{n}+1)^{-\frac{1}{2}}\hat{a}-\hat{a}^\dagger(\hat{n}+1)^{-\frac{1}{2}}\right],
\end{eqnarray}
\noindent
are quadratures of the readout observable. It is straightforward to show by induction that $[f(\hat{a}^\dagger),\hat{a}]=-f'(\hat{a}^\dagger)$ and $[\hat{a}^\dagger,f(\hat{a})]=-f'(\hat{a})$ with $f(\cdot):\mathcal{R}\mapsto\mathcal{R}$ being a real function of its argument. Now, the non-zero commutators of the basis $\{A\}^{\rm T}$ can be found after some algebra as $[\hat{n},\hat{C}]=-i\hat{S}$, $[\hat{n},\hat{S}]=i\hat{C}$, and $[\hat{C},\hat{S}]=\frac{1}{2}i(\hat{n}+2)^{-1}$. All remains to construct the Langevin equations now, is to linearize the last commutators as $[\hat{C},\hat{S}]\approx \frac{1}{2}i(\bar{n}+2)^{-1}$, by which the basis $\{A\}^{\rm T}=\{\hat{n},\hat{m},\hat{C},\hat{S}\}$ would satisfy closedness. Input noise terms to the operators $\hat{C}$ and $\hat{S}$ should be constructed by linear combinations of $\hat{a}_{\rm in}$ and $\hat{a}_{\rm in}^\dagger$ while replacing the multiplier term $1/\sqrt{\hat{n}+1}$ with the linearized form $1/\sqrt{\bar{n}+1}$.

In the end, it has to be mentioned that under external drive, periodicity, or dynamical control $[\textbf{M}(t)]$ in (\ref{eq6}) is time-dependent \cite{36f,51}. For instance, the ultimate optomechanical cooling limit is a function of system dynamics \cite{56}. Then, integration should be done numerically, since exact analytical solutions without infinite perturbations exist only for very restricted cases. This is, however, beyond the scope of the current study.



\section{Conclusions}

A new method was described to solve quadratic quantum interactions using perturbative truncation schemes, by including higher-order operators in the solution space. Spectral densities of higher-order operators, calculation of the second-order correlation function, as well as the quantum anharmonic oscillator and transformation to scalar forms using $Q-$functions were discussed. Finally, applications of the presented approach to quantum limited amplifiers, and nondemolition measurements were demonstrated.  

\acknowledgments{
	This work been supported by Laboratory of Photonics and Quantum Measurements at \'{E}cole Polytechnique F\'{e}d\'{e}rale de Lausanne and Research Deputy of Sharif University of Technology. The author thanks Prof. Franco Nori, Prof. Vincenzo Savona, Dr. Alexey Feofanov, Dr. Christophe Galland, Dr. Sahar Sahebdivan, as well as Liu Qiu and Amir H. Ghadimi for comments and/or discussions. The author is highly indebted to Dr. Hiwa Mahmoudi at Institute of Electrodynamics, Microwave and Circuit Engineering in Technische Universit\"{a}t Wien, and in particular, the Laboratory for Quantum Foundations and Quantum Information on the Nano- and Microscale in Vienna Center for Quantum Science and Technology (VCQ) at Universit\"{a}t Wien for their warm and receptive hospitality during which the numerical computations and final revisions took place. The huge effort needed in improving the presentation of this article has not been possible without support and encouragement of the celebrated artist, Anastasia Huppmann.
}

\appendix
\section*{Appendices}

We show by numerical solution of a stochastic nonlinear operator differential equation, and also taking its expectation values within the Mean Field Approximation, that the proposed analytical scheme in the above referenced manuscript works very well, is convergent, and uniformly converges to the accurate solution. 

We also demonstrate the second-order exact solution to the standard optomechanical Hamiltonian, and show that the proposed method can reproduce the side-band asymmetry in quantum optomechanics surprisingly well, which is a rigorous proof of the quantum mechanical capability of the proposed approach.

\section{Nonlinear First-order Circuit}

First, consider the infinitely-ordered nonlinear operator equation

\begin{equation}
\label{eqS1}
\tau \frac{d}{dt}\hat{u}\left(t\right)=-\mu \hat{u}\left(t\right)-\kappa \left[e^{\hat{u}\left(t\right)}-1\right]+v\left(t\right)-\hat{n}\left(t\right),
\end{equation}
\noindent 
which models the voltage operator of an RC circuit shunted by a nonlinear ideal diode, driven by a sinusoidal voltage source $v\left(t\right)=V_0e^{-\alpha t}\sin\left(\omega t\right)$, and stochastic noise $\hat{n}\left(t\right)$. We suppose that the noise $\hat{n}\left(t\right)$ is governed by a Weiner process. Here, and without loss of generality, both $\kappa $ and $\mu $ are taken to be positive real parameters. Hence, this model does not include an oscillating part due to an imaginary $\mu $, which could have been otherwise absorbed into $\hat{u}\left(t\right)$ by a rotating frame transformation. This particular choice also eliminates the imaginary part of $\hat{u}\left(t\right)$. We also assume here, for the illustrative purpose of this example, that $\mu =V_0=\tau =1$, $\omega =2\pi \alpha ,$ and $\omega =2\pi \times \text{1kHz}$.

The reason for choosing this particular differential equation is that it is nonlinear to the infinite order, and also the extended basis of higher order operators all commute and therefore trivially form a closed basis.

Using the proposed method in the paper under consideration, this above operator equation can be first put into the infinitely-ordered linear system of ordinary differential equations as

\begin{eqnarray}
\label{eqS2}
\tau \frac{d}{dt}\left\{ \begin{array}{c}
\hat{u}\left(t\right) \\ 
{\hat{u}}^2\left(t\right) \\ 
\begin{array}{c}
{\hat{u}}^3\left(t\right) \\ 
{\hat{u}}^4\left(t\right) \\ 
\begin{array}{c}
{\hat{u}}^5\left(t\right) \\ 
\vdots  \end{array}
\end{array}
\end{array}
\right\}=-\left[ \begin{array}{c}
\begin{array}{cccccc}
\kappa +1 & \frac{\kappa }{2!} & \frac{\kappa }{3!} & \frac{\kappa }{4!} & \frac{\kappa }{5!} & \cdots  \\ 
0 & 2\left(\kappa +1\right) & \frac{2\kappa }{2!} & \frac{2\kappa }{3!} & \frac{2\kappa }{4!} & \cdots  \\ 
0 & 0 & 3\left(\kappa +1\right) & \frac{3\kappa }{2!} & \frac{3\kappa }{3!} & \cdots  \\ 
0 & 0 & 0 & 4\left(\kappa +1\right) & \frac{4\kappa }{2!} & \cdots  \\ 
0 & 0 & 0 & 0 & 5\left(\kappa +1\right) & \cdots  \\ 
\vdots  & \vdots  & \vdots  & \vdots  & \vdots  & \ddots  \end{array}
\end{array}
\right]\left\{ \begin{array}{c}
\hat{u}\left(t\right) \\ 
{\hat{u}}^2\left(t\right) \\ 
\begin{array}{c}
{\hat{u}}^3\left(t\right) \\ 
{\hat{u}}^4\left(t\right) \\ 
\begin{array}{c}
{\hat{u}}^5\left(t\right) \\ 
\vdots  \end{array}
\end{array}
\end{array}
\right\}\\ \nonumber 
+\left\{ \begin{array}{c}
v\left(t\right) \\ 
2\hat{u}\left(t\right)v\left(t\right) \\ 
\begin{array}{c}
3{\hat{u}}^2\left(t\right)v\left(t\right) \\ 
4{\hat{u}}^3\left(t\right)v\left(t\right) \\ 
\begin{array}{c}
5{\hat{u}}^4\left(t\right)v\left(t\right) \\ 
\vdots  \end{array}
\end{array}
\end{array}
\right\}-\left\{ \begin{array}{c}
\hat{n}\left(t\right) \\ 
2\hat{u}\left(t\right)\hat{n}\left(t\right) \\ 
\begin{array}{c}
3{\hat{u}}^2\left(t\right)\hat{n}\left(t\right) \\ 
4{\hat{u}}^3\left(t\right)\hat{n}\left(t\right) \\ 
\begin{array}{c}
5{\hat{u}}^4\left(t\right)\hat{n}\left(t\right) \\ 
\vdots  \end{array}
\end{array}
\end{array}
\right\}.
\end{eqnarray}

Subsequently, the input terms can be linearized using the proposed method in the paper. Doing this results in 

\begin{eqnarray}
\label{eqS3}
\tau \frac{d}{dt}\left\{ \begin{array}{c}
\hat{u}\left(t\right) \\ 
{\hat{u}}^2\left(t\right) \\ 
\begin{array}{c}
{\hat{u}}^3\left(t\right) \\ 
{\hat{u}}^4\left(t\right) \\ 
\begin{array}{c}
{\hat{u}}^5\left(t\right) \\ 
\vdots  \end{array}
\end{array}
\end{array}
\right\}=-\left[ \begin{array}{c}
\begin{array}{cccccc}
\kappa +1 & \frac{\kappa }{2!} & \frac{\kappa }{3!} & \frac{\kappa }{4!} & \frac{\kappa }{5!} & \cdots  \\ 
0 & 2\left(\kappa +1\right) & \frac{2\kappa }{2!} & \frac{2\kappa }{3!} & \frac{2\kappa }{4!} & \cdots  \\ 
0 & 0 & 3\left(\kappa +1\right) & \frac{3\kappa }{2!} & \frac{3\kappa }{3!} & \cdots  \\ 
0 & 0 & 0 & 4\left(\kappa +1\right) & \frac{4\kappa }{2!} & \cdots  \\ 
0 & 0 & 0 & 0 & 5\left(\kappa +1\right) & \cdots  \\ 
\vdots  & \vdots  & \vdots  & \vdots  & \vdots  & \ddots  \end{array}
\end{array}
\right]\left\{ \begin{array}{c}
\hat{u}\left(t\right) \\ 
{\hat{u}}^2\left(t\right) \\ 
\begin{array}{c}
{\hat{u}}^3\left(t\right) \\ 
{\hat{u}}^4\left(t\right) \\ 
\begin{array}{c}
{\hat{u}}^5\left(t\right) \\ 
\vdots  \end{array}
\end{array}
\end{array}
\right\}\\ \nonumber
+\left\{ \begin{array}{c}
v\left(t\right) \\ 
2\bar{u}v\left(t\right) \\ 
\begin{array}{c}
3{\bar{u}}^2v\left(t\right) \\ 
4{\bar{u}}^3v\left(t\right) \\ 
\begin{array}{c}
5{\bar{u}}^4v\left(t\right) \\ 
\vdots  \end{array}
\end{array}
\end{array}
\right\}-\left\{ \begin{array}{c}
\hat{n}\left(t\right) \\ 
2\bar{u}\hat{n}\left(t\right) \\ 
\begin{array}{c}
3{\bar{u}}^2\hat{n}\left(t\right) \\ 
4{\bar{u}}^3\hat{n}\left(t\right) \\ 
\begin{array}{c}
5{\bar{u}}^4\hat{n}\left(t\right) \\ 
\vdots  \end{array}
\end{array}
\end{array}
\right\},
\end{eqnarray}
\noindent in which $\bar{u}=\frac{1}{T}\int^T_0{\left\langle \hat{u}\left(t\right)\right\rangle dt}$ is the time-average of the input. Now, the above system of equations can be exactly integrated, after truncation to a finite-order.

\begin{figure}
	\centering
	\includegraphics[width=4in]{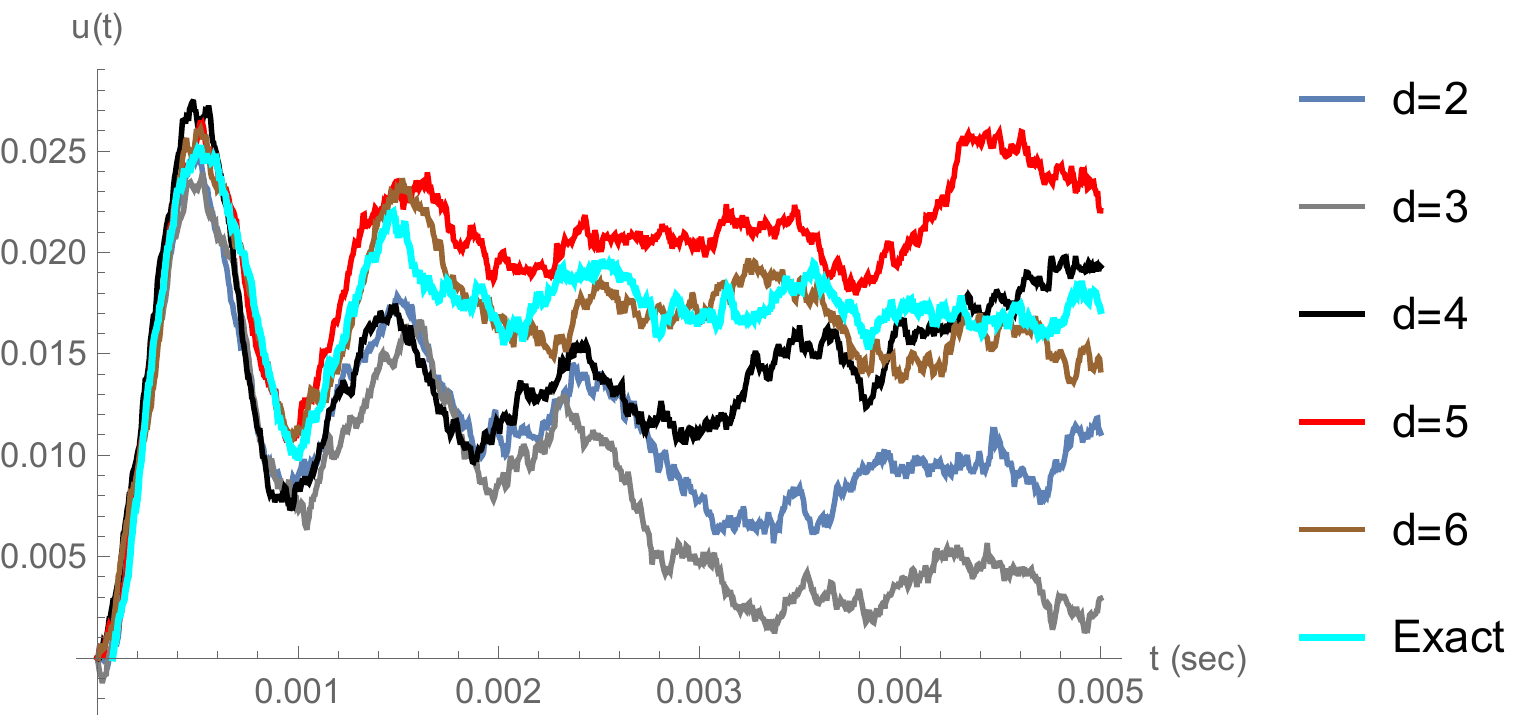}
	\caption{The stochastic solution function $u\left(t\right)=\left\langle \hat{u}\left(t\right)\right\rangle $ versus time given in various orders of approximation. \label{FigS1}}
\end{figure}

Using an extensive code written in Mathematica, the above system of linear stochastic equations can be treated and integrated as an It\^{o} process, and the results for various orders of truncation between 2 and 6 versus the numerically exact solution are displayed in Fig. \ref{FigS1}.

It is still not quite clear that the method is convergent to the exact solution, since the It\^{o} integration of a Weiner process every time is carried over a different sequence of random numbers. This difficulty cannot be avoided in principle, since there is no way to reset the numerically random sequence.

Therefore, as a double check, we take the expectation values, which discards the noise term, and transform a mean-field approximation to reach a similar system of differential equations, however, expressed in terms of the expectation value function $\left\langle \hat{u}\left(t\right)\right\rangle $ and its higher orders. This is equivalent to solving the nonlinear differential equation

\begin{equation}
\label{eqS4}
\tau \frac{d}{dt}\left\langle \hat{u}\left(t\right)\right\rangle =-\mu \left\langle \hat{u}\left(t\right)\right\rangle -\kappa \left[e^{\left\langle \hat{u}\left(t\right)\right\rangle }-1\right]+v\left(t\right),
\end{equation}
\noindent 
given the fact that $\left\langle \hat{n}\left(t\right)\right\rangle =0$.

\begin{figure}[!htb]
	\centering
	\includegraphics[width=4in]{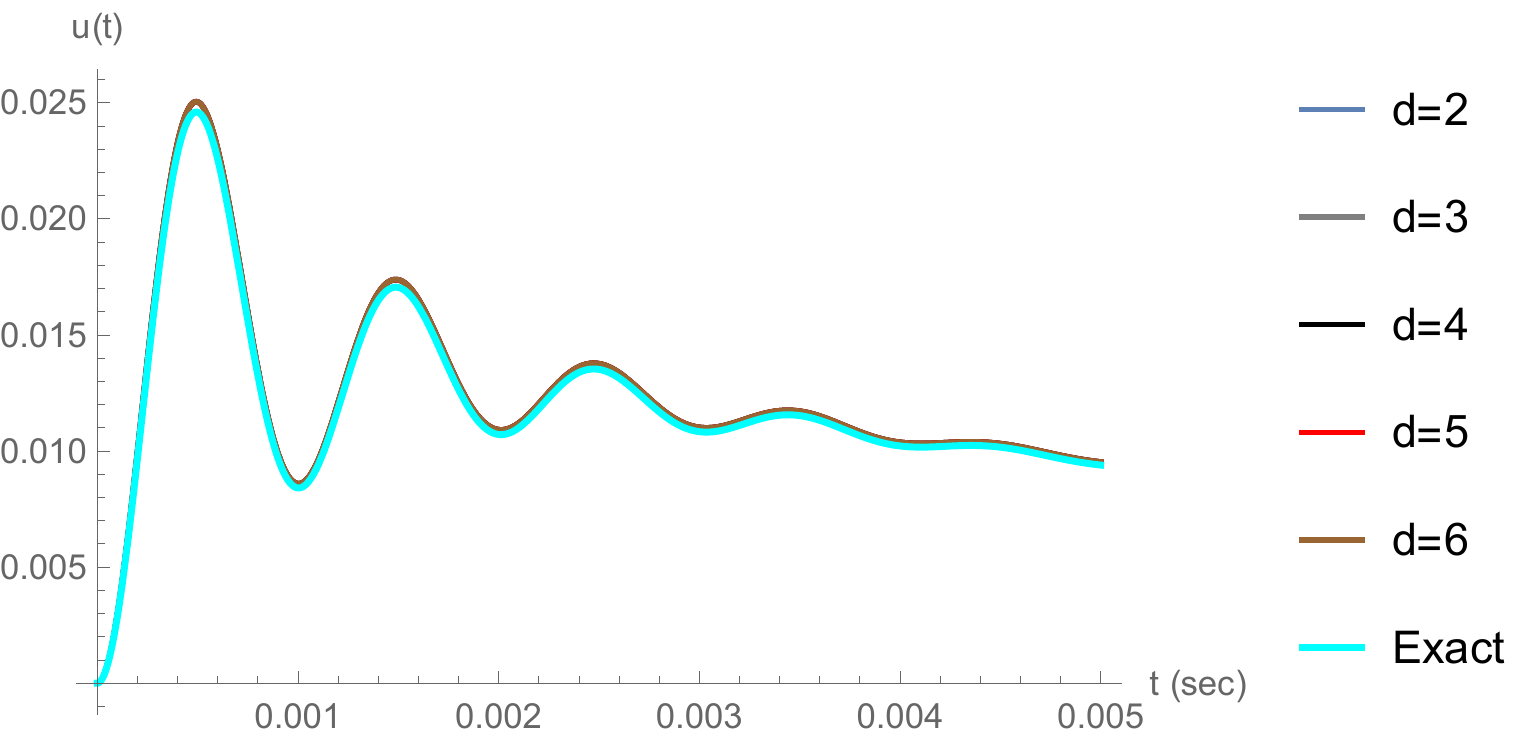}
	\caption{Expectation value function $\left\langle \hat{u}\left(t\right)\right\rangle $ versus time given in various orders of approximation. Convergence to the exact solution obtained from numerical solution of (\ref{eqS4}) is rapid by increasing order.\label{FigS2}}
\end{figure}

Doing this immediately reveals the convergence property of our proposed method, illustrated in Fig. \ref{FigS2}. As it can be clearly verified, the numerical solutions are so rapidly and accurately converging to the exact solution, that they are practically indistinguishable beyond the two lowest truncation orders.

\section{Optomechanical Hamiltonian}

The standard optomechanical Hamiltonian is not quadratic, but can be solved within the second-order accuracy using the method described in the article. Here, we are able to show that the proposed approach actually can reproduce fundamental quantum mechanical properties of a typical nonlinear system, such as side-band asymmetry.

Its standard optomechanical Hamiltonian reads
\begin{equation}
\label{eqS5}
\mathbb{H}_{\text{OM}}=\hbar \Omega\hat{m}-\hbar \Delta\hat{n}-\hbar g_0\hat{n}\left(\hat{b}+{\hat{b}}^{\dagger }\right).
\end{equation}
\noindent
In order to form a closed basis of operators, we may choose
\begin{equation}
\label{eqS6}
A^\text{T}=\left\{\hat{a},\hat{b},\hat{a}\hat{b},\hat{a}{\hat{b}}^{\dagger },\hat{n},\hat{c}\right\},
\end{equation}
\noindent 
which forms a $6\times 6$ system of Langevin equations. It is easy to verify that this system is exactly closed, by calculation of all possible commutation pairs between the elements. Out of the $6!$ commutators, the non-zero ones are

\begin{eqnarray}
\label{eqS7}
\left[\hat{a},\hat{n}\right]&=&-\left[\hat{a}{\hat{b}}^{\dagger },\hat{b}\right]=\hat{a},\\ \nonumber
\left[\hat{a}\hat{b},\hat{n}\right]&=&\hat{a}\hat{b},\\ \nonumber 
\left[\hat{a}{\hat{b}}^{\dagger },\hat{n}\right]&=&\hat{a}{\hat{b}}^{\dagger },\\ \nonumber
\left[\hat{a}\hat{b},\hat{a}{\hat{b}}^{\dagger }\right]&=&\left[\hat{c},\hat{n}\right]=2\hat{c}, 
\end{eqnarray}
\noindent 
This is obviously a closed basis. Now, one may proceed with composition of the Langevin equations. They are given by 

\begin{eqnarray}
\label{eqS8}
\frac{d}{dt}\left\{ \begin{array}{c}
\hat{a} \\ 
\hat{b} \\ 
\begin{array}{c}
\hat{a}\hat{b} \\ 
\hat{a}{\hat{b}}^{\dagger } \\ 
\begin{array}{c}
\hat{n} \\ 
\hat{c} \end{array}
\end{array}
\end{array}
\right\}=\left[ \begin{array}{c}
\begin{array}{cccccc}
-i\Delta-\frac{\kappa }{2} & 0 & ig_0 & ig_0 & 0 & 0 \\ 
0 & i\Omega-\frac{\Gamma}{2} & 0 & 0 & ig_0 & 0 \\ 
ig_0\left(\hat{n}+\hat{m}+1\right) & 0 & i\left(\Omega-\Delta\right)-\frac{\gamma }{2} & ig_0 & 0 & 0 \\ 
ig_0\left(\hat{n}+\hat{m}+1\right) & 0 & ig_0 & -i\left(\Omega+\Delta\right)-\frac{\gamma }{2} & 0 & 0 \\ 
0 & 0 & 0 & 0 & -\kappa  & 0 \\ 
0 & 0 & ig_0\hat{a} & ig_0\hat{a} & 0 & -2i\Delta-\kappa  \end{array}
\end{array}
\right]\left\{ \begin{array}{c}
\hat{a} \\ 
\hat{b} \\ 
\begin{array}{c}
\hat{a}\hat{b} \\ 
\hat{a}{\hat{b}}^{\dagger } \\ 
\begin{array}{c}
\hat{n} \\ 
\hat{c} \end{array}
\end{array}
\end{array}
\right\}\\ \nonumber 
+\left\{ \begin{array}{c}
\sqrt{\kappa }{\hat{a}}_{\text{in}} \\ 
\sqrt{\Gamma}{\hat{b}}_{\text{in}} \\ 
\begin{array}{c}
\sqrt{\gamma }{\left(\hat{a}\hat{b}\right)}_{\text{in}} \\ 
\sqrt{\gamma }{\left(\hat{a}{\hat{b}}^{\dagger }\right)}_{\text{in}} \\ 
\begin{array}{c}
\sqrt{\kappa }{\hat{n}}_{\text{in}} \\ 
\sqrt{\kappa }{\hat{c}}_{\text{in}} \end{array}
\end{array}
\end{array}
\right\},
\end{eqnarray}
\noindent 
where $\gamma =\kappa +\Gamma$,  we have set $\hat{x}=\hat{a}$ in (10) in all equations, ${\hat{n}}_\text{in}={\hat{a}}^{\dagger }{\hat{a}}_\text{in}+\hat{a}{{\hat{a}}^{\dagger }}_\text{in}$ and ${\hat{c}}_\text{in}=2\hat{a}{\hat{a}}_\text{in}$. This subsequently can be linearized to get the second-order accurate optomechanical system of equations as

\begin{eqnarray}
\label{eqS9}
\frac{d}{dt}\left\{ \begin{array}{c}
\hat{a} \\ 
\hat{b} \\ 
\begin{array}{c}
\hat{a}\hat{b} \\ 
\hat{a}{\hat{b}}^{\dagger } \\ 
\begin{array}{c}
\hat{n} \\ 
\hat{c} \end{array}
\end{array}
\end{array}
\right\}=\left[ \begin{array}{c}
\begin{array}{cccccc}
-i\Delta-\frac{\kappa }{2} & 0 & ig_0 & ig_0 & 0 & 0 \\ 
0 & i\Omega-\frac{\Gamma}{2} & 0 & 0 & ig_0 & 0 \\ 
iL & 0 & i\left(\Omega-\Delta\right)-\frac{\gamma }{2} & ig_0 & 0 & 0 \\ 
iL & 0 & ig_0 & -i\left(\Omega+\Delta\right)-\frac{\gamma }{2} & 0 & 0 \\ 
0 & 0 & 0 & 0 & -\kappa  & 0 \\ 
0 & 0 & i2F & i2F & 0 & -2i\Delta-\kappa  \end{array}
\end{array}
\right]\left\{ \begin{array}{c}
\hat{a} \\ 
\hat{b} \\ 
\begin{array}{c}
\hat{a}\hat{b} \\ 
\hat{a}{\hat{b}}^{\dagger } \\ 
\begin{array}{c}
\hat{n} \\ 
\hat{c} \end{array}
\end{array}
\end{array}
\right\}\\ \nonumber
+\left\{ \begin{array}{c}
\sqrt{\kappa }{\hat{a}}_{\text{in}} \\ 
\sqrt{{\Gamma}}{\hat{b}}_{\text{in}} \\ 
\begin{array}{c}
\sqrt{\gamma }{\left(\hat{a}\hat{b}\right)}_{\text{in}} \\ 
\sqrt{\gamma }{\left(\hat{a}{\hat{b}}^{\dagger }\right)}_{\text{in}} \\ 
\begin{array}{c}
\sqrt{K}{\hat{a}}_{\text{in}} \\ 
\sqrt{2}K{\hat{a}}_{\text{in}}+\sqrt{8\kappa }{{\hat{a}}_{\text{in}}}^2 
\end{array}
\end{array}
\end{array}
\right\},
\end{eqnarray}
\noindent 
in which $F=g_0\sqrt{\bar{n}}$, $K=4\bar{n}\kappa $, $G=g_0\bar{n}$, $L=G+g_0\left(\bar{m}+1\right)$, and can be further approximated by $L=G$ under normal experimental conditions of an ultracold cavity. The average population value $\bar{m}={1}/{\left[\exp\left({\hbar \Omega}/{k_\text{B}T}\right)\right]}$, while $\bar{n}$ can be obtained from the steady state solution of the first row by replacements of input noise term $\sqrt{\kappa}{\hat{a}}_\text{in}\to \alpha$, where $\alpha $ is the input photon flux. 

\begin{figure}[!htb]
	\centering
	\includegraphics[width=4in]{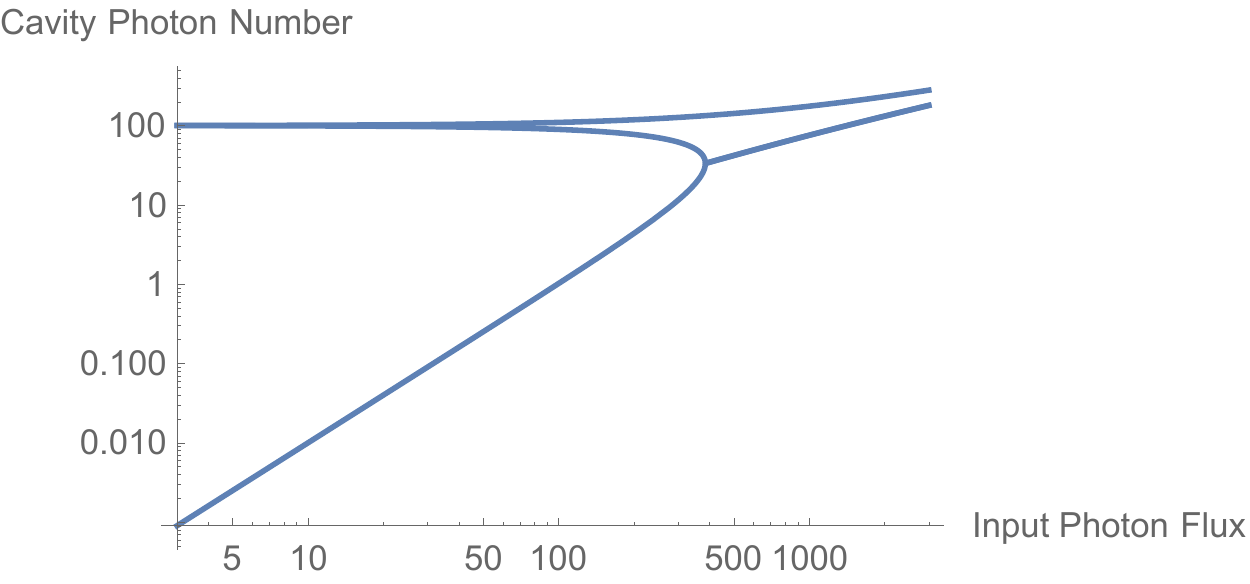}
	\caption{Bistable behavior of cavity photon number $\bar{n}$ versus input photon flux $\alpha $.\label{FigS3}}
\end{figure}

\begin{figure}[!htb]
	\centering
	\includegraphics[width=4in]{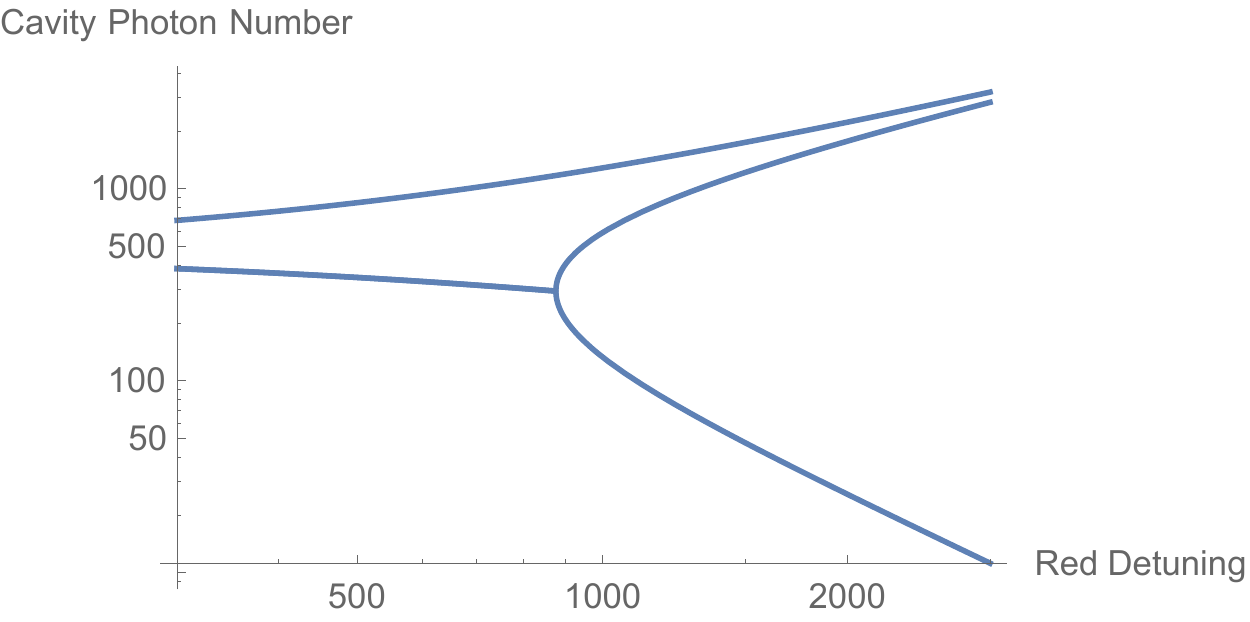}
	\caption{Bistable behavior of cavity photon number $\bar{n}$ versus red detuning $\Delta$.\label{FigS4}}
\end{figure}

Doing this gives the third order equation in terms of $\sqrt{\bar{n}}$ as

\begin{eqnarray}
\label{eqS10a}
iL\sqrt{\bar{n}}&+&\left[i\left(g_0+\Omega-\Delta\right)-\frac{\gamma }{2}\right]\sqrt{\bar{n}\bar{m}}\\ \nonumber
&=&ig_0\left({\sqrt{\bar{n}}}^2+\bar{m}+1\right)\sqrt{\bar{n}}+\sqrt{\bar{m}}\left[i\left(g_0+\Omega-\Delta\right)-\frac{\gamma }{2}\right]\sqrt{\bar{n}} \\ \nonumber
&=&ig_0{\sqrt{\bar{n}}}^3+\left[ig_0\left(\bar{m}+1\right)\sqrt{\bar{m}}+i\left(g_0+\Omega-\Delta\right)-\frac{\gamma }{2}\right]\sqrt{\bar{n}}=\sqrt{\bar{m}}\alpha \\ \nonumber 
&=&ig_0{\sqrt{\bar{n}}}^3+iB\sqrt{\bar{n}}=C,\\ 
\label{eqS10b}
A&=&g_0,\\ \nonumber 
B&=&g_0\left(\bar{m}+1\right)\sqrt{\bar{m}}+\left(g_0+\Omega-\Delta\right)+i\frac{\gamma }{2}=B_r+i\frac{\gamma }{2},\\ \nonumber
C&=&\sqrt{\bar{m}}\alpha.
\end{eqnarray}
\noindent 
Solution of this third-order algebraic equation gives the three distinct solutions

\begin{eqnarray}
\label{eqS11}
{\left.\bar{n}\right|}_1&=&\frac{1}{3\sqrt[3]{12}g_0^2 |Z|^2}\left|\sqrt[3]{12}g_0 B+Z^2\right|^2,\\ \nonumber
{\left.\bar{n}\right|}_2&=&\frac{1}{12\sqrt[3]{12}g_0^2 |Z|^2} \left|\sqrt[3]{12}(1+i\sqrt{3})g_0 B+(1-i\sqrt{3})Z^2\right|^2,\\ \nonumber 
{\left.\bar{n}\right|}_3&=&\frac{1}{12\sqrt[3]{12}g_0^2 |Z|^2} \left|\sqrt[3]{12}(1-i\sqrt{3})g_0 B+(1+i\sqrt{3})Z^2\right|^2,
\end{eqnarray}
\noindent 
in which
\begin{equation}
Z=\sqrt[3]{9{g_0}^2C\left(1-\sqrt{1-\frac{4B^3}{27g_0C^2}}\right)},
\end{equation} 
is a complex number. These three roots clearly become two distinct solutions in the limit of lossless cavity with $\gamma =0$ as long as $27g_0C^2>4B_r^2$, and very well reproduce the expected bistable behavior of cavity photon number $\bar{n}$ with respect to various parameters such as detuning $\Delta$ and input photon flux $\alpha $. This has been shown for typical normalized variables in Figs. \ref{FigS3} and \ref{FigS4}.

For the noise terms, we may insert 

\begin{eqnarray}
\label{eqS12}
\sqrt{\gamma }{\left(\hat{a}\hat{b}\right)}_\text{in}&=&{\sqrt{\gamma }\left(\hat{a}{\hat{b}}^{\dagger }\right)}_\text{in}=\sqrt{\Gamma}\hat{a}{\hat{b}}_\text{in}+\sqrt{\kappa}{\hat{a}}_\text{in}\hat{b}\to \sqrt{\Gamma\alpha}{\hat{b}}_\text{in}+\sqrt{\kappa \bar{m}}{\hat{a}}_\text{in},\\ \nonumber
\sqrt{\kappa }{\hat{n}}_\text{in}&=&\sqrt{\kappa }{\hat{a}}^{\dagger }{\hat{a}}_\text{in}+\sqrt{\kappa }\hat{a}{{\hat{a}}^{\dagger }}_\text{in}\to \sqrt{K}{\hat{a}}_\text{in},\\ \nonumber
{\hat{c}}_{in}&=&2\hat{a}{\hat{a}}_\text{in}\to 2\sqrt{\bar{n}}{\hat{a}}_\text{in}+2{{\hat{a}}_\text{in}}^2.
\end{eqnarray}
\noindent 
Here, the spectral density of ${{\hat{a}}_\text{in}}^2$ has already been calculated in Section \ref{NoiseSpectra}. As opposed to the second-order accurate optomechanical Langevin equations, the first-order accurate equations are simply obtained by truncating (\ref{eqS8}) which recovers the well-known $3\times 3$ system with the obviously closed basis $A^\text{T}=\left\{\hat{a},\hat{b},{\hat{b}}^{\dagger }\right\}$ after simple algebraic manipulation

\begin{equation}
\label{eqS13}
\frac{d}{dt}\left\{ \begin{array}{c}
\hat{a} \\ 
\hat{b} \\ 
{\hat{b}}^{\dagger } \end{array}
\right\}=\left[ \begin{array}{ccc}
-i\Delta-\frac{\kappa }{2} & iF & iF \\ 
0 & i\Omega-\frac{\Gamma}{2} & 0 \\ 
0 & 0 & -i\Omega-\frac{\Gamma}{2} \end{array}
\right]\left\{ \begin{array}{c}
\hat{a} \\ 
\hat{b} \\ 
{\hat{b}}^{\dagger } \end{array}
\right\}+\left\{ \begin{array}{c}
0 \\ 
iG \\ 
-iG \end{array}
\right\}+\left\{ \begin{array}{c}
\sqrt{\kappa }{\hat{a}}_\text{in} \\ 
\sqrt{\Gamma}{\hat{b}}_\text{in} \\ 
\sqrt{\Gamma}{{\hat{b}}_\text{in}}^{\dagger } \end{array}
\right\}.
\end{equation}

Here, we have used the further replacements $g_0\hat{n}\to G$ and $g_0\hat{a}\hat{b}\to F\hat{b}$. However, if we had made the replacement $g_0\hat{n}\to F\hat{a}$ we would get

\begin{equation}
\label{eqS14}
\frac{d}{dt}\left\{ \begin{array}{c}
\hat{a} \\ 
\hat{b} \\ 
{\hat{b}}^{\dagger } \end{array}
\right\}=\left[ \begin{array}{ccc}
-i\Delta-\frac{\kappa }{2} & iF & iF \\ 
iF & i\Omega-\frac{\Gamma}{2} & 0 \\ 
0 & iF & -i\Omega-\frac{\Gamma}{2} \end{array}
\right]\left\{ \begin{array}{c}
\hat{a} \\ 
\hat{b} \\ 
{\hat{b}}^{\dagger } \end{array}
\right\}+\left\{ \begin{array}{c}
\sqrt{\kappa }{\hat{a}}_\text{in} \\ 
\sqrt{\Gamma}{\hat{b}}_\text{in} \\ 
\sqrt{\Gamma}{{\hat{b}}_\text{in}}^{\dagger } \end{array}
\right\}.
\end{equation}

Finally, we are able to show that this proposed approach can reproduce the side-band asymmetry of an optomechanical cavity, which is a fundamental quantum mechanical property. 

In the context of quantum optomechanics, it is general practice to solve either (\ref{eqS13}) or (\ref{eqS14}) with \textit{asymmetric} noise spectral densities such as $S_{BB}(+\omega)\neq S_{BB}(-\omega)$. While being a purely nonlinear quantum mechanical effect, this is the only known way at this moment to reproduce the side-band asymmetry. Here, we can show indeed that asymmetric spectral noise terms are \textit{not needed} to obtain the much expected side-band asymmetry.

This result becomes possible by considering that in the system (\ref{eqS9}) both of the terms $\hat{a}\hat{b}$ and $\hat{a}\hat{b}^\dagger$ are present in the second-order accurate basis. The first term and its conjugate $\{\hat{a}\hat{b},\hat{a}^\dagger\hat{b}^\dagger\}$ represent the parametric interactions on the blue side, while the second term and its conjugate $\{\hat{a}\hat{b}^\dagger,\hat{a}\hat{b}^\dagger\}$ represent the hopping interactions on the red side. By assumption of some test values to the input parameters as well as \textit{symmetric} input noise terms such as $S_{BB}(+\omega)=S_{BB}(-\omega)$, the spectral density of the collected output light from the cavity can be calculated. 

\begin{figure}[!htb]
	\centering
	\includegraphics[width=4in]{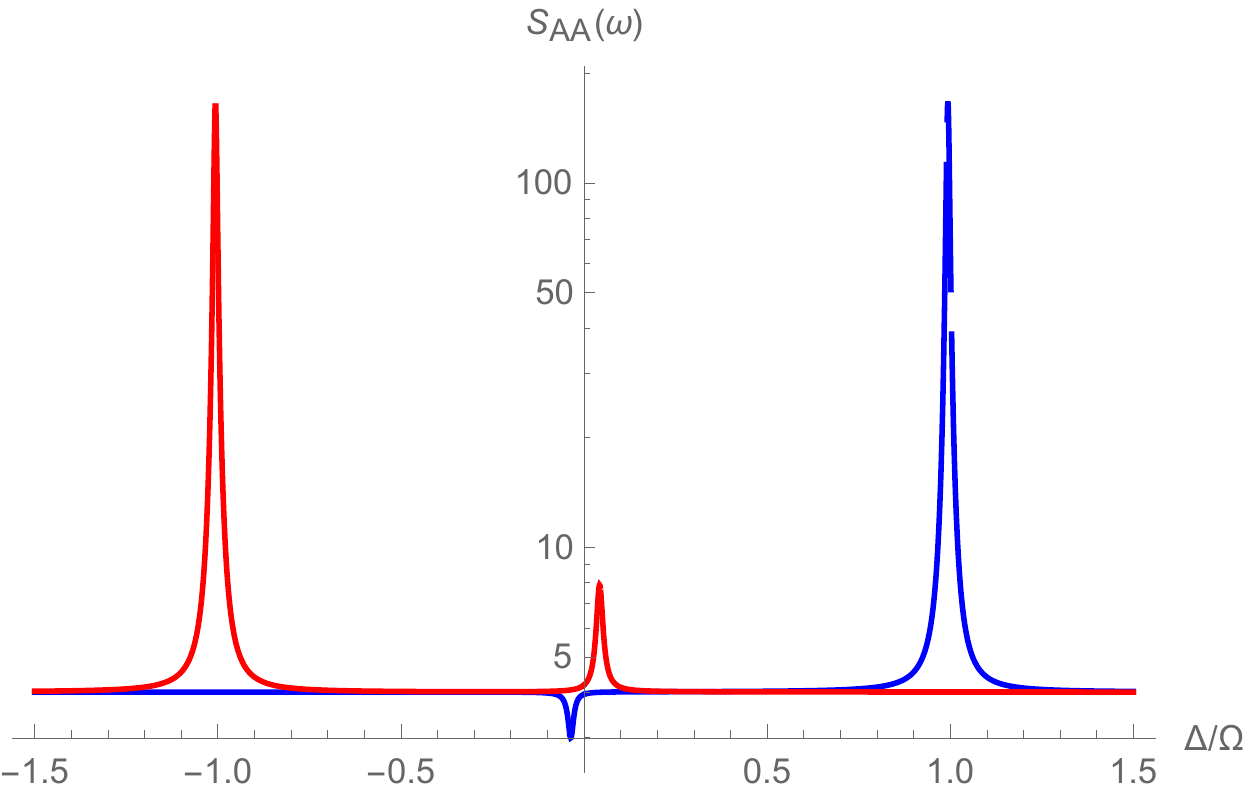}
	\caption{Spectral densities of an optomechanical cavity from solution of (\ref{eqS9}) with \textit{symmetric} input noise terms. Red and blue curves correspond to the pumping on red and blue sides. Horizontal axis is the normalized detuning with respect to the mechanical frequency. \label{FigS5}}
\end{figure}

\begin{figure}[!htb]
	\centering
	\includegraphics[width=4in]{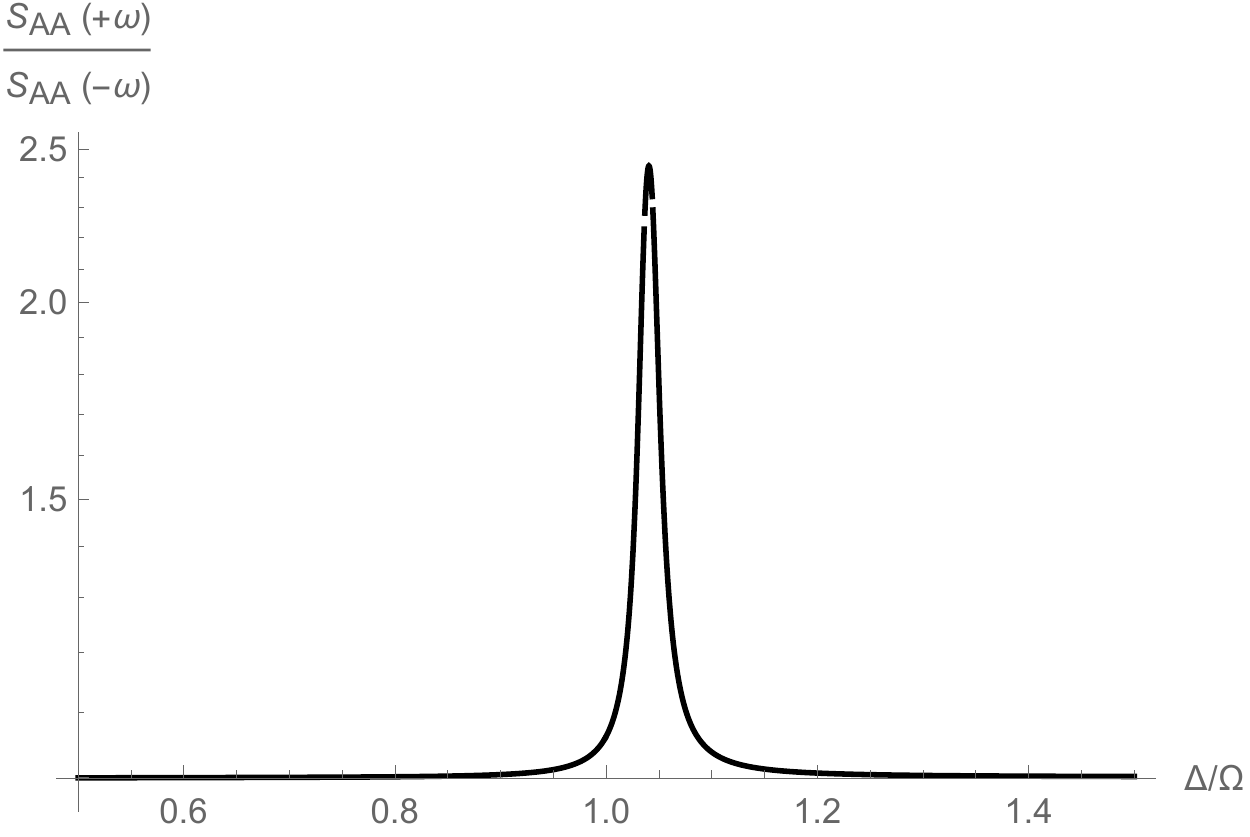}
	\caption{Side-band asymmetry, defined as the ratio of spectral densities at opposite frequencies, in an optomechanical cavity obtained from numerical solution of (\ref{eqS9}) with \textit{symmetric} input noise terms. Horizontal axis is the normalized detuning with respect to the mechanical frequency. \label{FigS6}}
\end{figure}

This has been done for the two cases of pumping on the red-side and blue-side shown in Fig. \ref{FigS5}. More interestingly and as the ultimate verification of the approach, the pumping can be tried exactly on the cavity resonance. Carrying out this procedure allows calculation of cavity sidebands at blue and red, the ratio of which is here plotted and shown in Fig. \ref{FigS6}. This figure reveals that there actually exists an easily observable asymmetry, which is hallmark of nonlinear quantum mechanical effects in cavity optomechanics. This should have rigorously proven, while leaving no further doubt, that the second-order accurate system of equations (\ref{eqS9}) can reproduce the side-band asymmetry in a satisfactory manner.

Detailed numerical simulation of standard and non-standard optomechanical problems using the proposed approach in paper, and verification against experiments remains as a subject of a future study.

\end{document}